\documentclass[article,superscriptaddress,11pt]{revtex4-1}

\usepackage{graphicx}
\usepackage{amsmath}
\usepackage{amssymb}
\usepackage{dcolumn}
\usepackage{bm}
\usepackage{slashed}
\usepackage{color}

\newcommand{\be}{\begin{equation}}
\newcommand{\ee}{\end{equation}}
\newcommand{\ba}{\begin{eqnarray}}
\newcommand{\ea}{\end{eqnarray}}
\newcommand{\no}{\nonumber \\}
\newcommand{\gsim}{\mathrel{\hbox{\rlap{\lower.55ex \hbox {$\sim$}}
                   \kern-.3em \raise.4ex \hbox{$>$}}}}
\newcommand{\lsim}{\mathrel{\hbox{\rlap{\lower.55ex \hbox {$\sim$}}
                   \kern-.3em \raise.4ex \hbox{$<$}}}}

\def\roughly#1{\mathrel{\raise.3ex\hbox{$#1$\kern-.75em%
\lower1ex\hbox{$\sim$}}}}
\def\lsim{\roughly<}
\def\gsim{\roughly>}

\def\({\left(}
\def\){\right)}
\def\[{\left[}
\def\]{\right]}
\def\<{\langle}
\def\>{\rangle}

\def\bB{{\bf B}}
\def\bna{{\boldsymbol\nabla}}
\def\bo{{\boldsymbol\omega}}
\def\tr{\text{tr}}
\def\tW{{\tilde W}}
\def\bW{{\bar W}}
\def\tP{{\d\P}_u}
\def\bP{{\d\P}_T}
\def\hP{{\d\P}}
\def\tj{{\tilde j}}

\def\cN{{\cal N}}
\def\cJ{{\cal J}}
\def\cE{{\cal E}}
\def\cP{{\cal P}}
\def\cQ{{\cal Q}}
\def\cT{{\cal T}}
\def\cA{{\cal A}}
\def\cC{{\cal C}}
\def\cD{{\cal D}}
\def\cV{{\cal V}}
\def\cF{{\cal F}}

\def\l{{\lambda}}
\def\L{{\Lambda}}
\def\d{{\delta}}
\def\D{{\Delta}}
\def\o{{\omega}}

\def\e{{\epsilon}}

\def\a{{\alpha}}
\def\b{{\beta}}
\def\c{{\chi}}

\def\p{{\pi}}
\def\P{{\Pi}}
\def\m{{\mu}}
\def\n{{\nu}}
\def\r{{\rho}}
\def\s{{\sigma}}

\def\th{{\theta}}

\def\x{{\xi}}
\def\P{{\Pi}}

\def\hb{{\hbar}}
\def\dg{{\dagger}}

\newcommand{\pd}{{\partial}}

\date{\today}

\begin{document}

\title{\bf Magneto-vortical Effect in Strong Magnetic Field}

\author{Shu Lin}
\email{linshu8@mail.sysu.edu.cn}
\affiliation{School of Physics and Astronomy, Sun Yat-Sen University, Zhuhai 519082, China}
\author{Lixin Yang}
\email{yanglx5@mail2.sysu.edu.cn}
\affiliation{School of Physics and Astronomy, Sun Yat-Sen University, Zhuhai 519082, China}

\begin{abstract}
  We develop  covariant chiral kinetic theory with Landau level basis. We use it to investigate a magnetized plasma with a transverse electric field and a steady vorticity as perturbations. After taking into account vacuum shift in the latter case, we find the resulting current and stress tensor in both cases can be matched consistently with constitutive equations of magnetohydrodynamics. We find the solution in the vorticity case contains both shifts in temperature and chemical potential as well as excitations of the lowest Landau level states. The solution gives rise to an vector charge density and axial current density. The vacuum parts coming from both shifts and excitations agree with previous studies and the medium parts coming entirely from excitations leads to a new contribution to vector charge and axial current density consistent with standard chiral vortical effect.
\end{abstract}

\maketitle


\newpage

\section{Introduction}

The response of QCD matter to magnetic field and vorticity has received much attention recently. In the linear regime, the response is the celebrated chiral magnetic effect (CME) \cite{Vilenkin:1980fu,Kharzeev:2004ey,Kharzeev:2007tn,Fukushima:2008xe,Son:2009tf,Neiman:2010zi} and chiral vortical effect (CVE) \cite{Vilenkin:1980zv,Erdmenger:2008rm,Banerjee:2008th,Son:2009tf,Neiman:2010zi,Landsteiner:2011cp}, which are known to be dictated by chiral anomaly and gravitational anomaly. While magnetic field and rotation are analogous in many ways, they differ in one crucial aspect. The magnetic field is external, but rotation is defined by motion of medium itself.

Recently the combined effect of magnetic field and vorticity has been studied by different groups \cite{Hattori:2016njk,Liu:2017spl,Chen:2015hfc,Cao:2019ctl,Chen:2019tcp,Bu:2019qmd,Fukushima:2020ncb}. In particular, it has been proposed by Hattori and Yin that in the limit of strong magnetic field, where lowest Landau level (LLL) approximation is valid, the effect of vorticity is to shift the energy of the LLL states through spin-orbit coupling \cite{Hattori:2016njk}
\begin{align}\label{E-shift}
  \D\e^{\pm}=\mp\frac{1}{2}\mathrm{sgn}(q_f)\hat{\bB}\cdot{\bo},
\end{align}
where the upper and lower signs correspond to particle and anti-particle of both chiralities and $q_f$ is the charge of particle. The energy shift can also be interpreted as a shift of chemical potential $\mathrm{sgn}(q_f)\hat{\bB}\cdot{\bo}$ for particle.
The shift induces vector charge density and axial current as
\begin{align}\label{mve}
  \D J_V^0=q_f\frac{1}{4\p^2}{\bB}\cdot{\bo},\quad \D {\bf J}_A=|q_f|\frac{1}{4\p^2}({\bB}\cdot{\bo})\hat{\bB}.
\end{align}
As remarked before, vorticity also implies circular motion of fluid velocity, which arises from average velocity of constituents in fluid cells. The rotation modifies the distribution of constituents in the plane transverse to the vorticity. It induces an extra contribution to \eqref{mve}. We will refer to this contribution as medium contribution, and \eqref{mve} as vacuum contribution based on their different origins.

Indeed, medium contributions to \eqref{mve} are expected. On the one hand, it is known that vector charge density receives the following contribution \cite{Kovtun:2016lfw}
\begin{align}\label{n_bound}
  \D J_V^0=-\bna \cdot{\bf P}-2{\bf M}\cdot{\bo}.
\end{align}
The first term is the familiar bound charge from polarization ${\bf P}$, which is absent in fluid. The second term is a required relativistic counterpart of the first one. It is from the coupling of magnetization ${\bf M}$ and vorticity. On the other hand, if we view $\D {\bf J}_A$ in \eqref{mve} as response to vorticity, we would expect also the standard CVE
\begin{align}\label{CVE_standard}
  \D {\bf J}_A=\(\frac{\m^2+\m_5^2}{2\p^2}+\frac{T^2}{6}\)\bo.
\end{align}

We will confirm the medium contributions in \eqref{n_bound} and \eqref{CVE_standard} in a magnetized quantum electrodynamics plasma with a vorticity. The combined vacuum and medium contributions can be matched nicely with constitutive equation of magnetohydrodynamics (MHD)\cite{Hernandez:2017mch,Grozdanov:2016tdf,Hongo:2020qpv,Hattori:2017usa} (see also \cite{Huang:2011dc,Finazzo:2016mhm}). We also study a closely related setting in which magnetized plasma is subject to transverse electric field. We will find the matching with MHD in this setting gives the same coefficients once the shift of chemical potential is carefully taken into account. In line with \eqref{mve}, we will work in the limit of strong magnetic field and use LLL approximation. The constituents of the fluid is LLL states and is described by chiral kinetic theory (CKT) with Landau level basis \cite{Lin:2019fqo,Hattori:2016lqx,Sheng:2017lfu}. It is supplementary to the usual chiral kinetic theory with free fermion basis, which is best suited at weak external field \cite{Son:2012wh,Son:2012zy,Stephanov:2012ki,Gao:2012ix,Pu:2010as,Chen:2012ca,Hidaka:2016yjf,Manuel:2013zaa,Manuel:2014dza,Wu:2016dam,Mueller:2017arw,Mueller:2017lzw,Huang:2018wdl,Gao:2018wmr,Carignano:2018gqt,Lin:2019ytz,Carignano:2019zsh,Liu:2018xip,Weickgenannt:2019dks,Gao:2019znl,Hattori:2019ahi,Wang:2019moi,Yang:2020hri,Liu:2020flb,Hayata:2020sqz,chen2021equaltime}.

This paper is organized as follows. In Section \eqref{sec_ckt}, we derive the CKT with Landau level basis in a covariant form. In the presence of strong magnetic field and vorticity in the fluid, the CKT needs to be corrected by the second order gradient terms on gauge potential. In Section \eqref{sec_drift}, we find solution for magnetized plasma subject to transverse electric field. The resulting drift correction to the current and stress tensor are matched with MHD. We then follow a similar procedure to obtain the solution for magnetized plasma with a vorticity and compare it with MHD in Section \eqref{sec_vortical}. We summarize and discuss possible extensions in Section \eqref{sec_summary}.

Throughout this paper, we set $\hb=1$ and $c=1$. We take positive charge $q_f=e$ for chiral fermions and absorb electric charge $e$ into the gauge field. We use the notations $(x^\m)=(x_0,{\bf x}),\,(p^\m)=(p_0,{\bf p})$ for four-vectors and adopt mostly minus signature.

\section{Covariant chiral kinetic theory with Landau level basis}\label{sec_ckt}

We start with a system of right-handed chiral fermions covariantly coupled to external gauge field. The two-point correlator $\bW(z,y)\equiv\<\psi(z)\psi(y)^\dg\>$ satisfies the following equations
\begin{align}\label{eom_bW}
  {\slashed D}_z\bW(z,y)=0,\quad \bW(z,y){\slashed D}_y^\dg=0,
\end{align}
with the covariant derivatives defined as
\begin{align}\label{cov_der}
  &{\slashed D}_z={\slashed \pd}_z+i{\slashed A}(z),\qquad{\slashed D}_y^\dg=\overleftarrow{{\slashed \pd}}_y-i{\slashed A}(y),
\end{align}
where, for right-handed fermions, the slash is given by ${\slashed A}=\s^\m A_\m$.
Note $\bW(z,y)$ is not gauge invariant. A gauge invariant correlator $\tW(z,y)$ is constructed by using a gauge link $U(y,z)$ as
\begin{align}\label{tw}
  \tW(z,y)\equiv \bW(z,y)U(y,z),
\end{align}
with the gauge link defined by
\begin{align}\label{link}
  U(y,z)=\mathrm{exp}\(i\int_z^y dr^\m A_\m(r)\).
\end{align}
In terms of $\tW(z,y)$, the EOM reads
\begin{align}\label{eom_tW}
  &{\slashed D}_z\bW(z,y)=D_\m^z\(\s^\m \tW(z,y)U(z,y)\)=0,\no
  &\bW(z,y){\slashed D}_y^\dg=\(\tW(z,y)U(z,y)\s^\m\)D_\m^{y\dg}=0.
\end{align}
It is convenient to switch to variables $x=\frac{1}{2}(y+z)$ and $s=y-z$. We will consider $x$ as a slow-varying variable and $s$ as a fast variable conjugate to momentum, i.e. $\pd_x\ll\pd_s$. This allows us to further simplify \eqref{eom_tW} using an expansion in $\pd_x$, which for the covariant derivatives and gauge link reads
\begin{align}\label{grad_exp}
  &D_\m^z=\frac{1}{2}\pd_\m^x+\pd_\m^s+iA_\m(x)+\frac{i}{2}\(s^\n \pd_\n^x\)A_\m(x)+\frac{i}{8}\(s^\n \pd_\n^x\)^2 A_\m(x)+O\((\pd_x)^3\), \no
  &D_\m^{y\dg}=\frac{1}{2}\overleftarrow{\pd}_\m^x-\overleftarrow{\pd}_\m^s-iA_\m(x)+\frac{i}{2}\(s^\n \pd_\n^x\)A_\m(x)-\frac{i}{8}\(s^\n \pd_\n^x\)^2 A_\m(x)+O\((\pd_x)^3\), \no
  &U(y,z)=\mathrm{exp}\(is^\m A_\m(x)+\frac{is^\m}{24}(s^\n \pd_\n^x)^2 A_\m(x)\)+O\((\pd_x)^4\).
\end{align}
Commuting the covariant derivatives with the gauge link using the following identities
\begin{align}
  &D_\m^z\(U(z,y)\s^\m \tW(z,y)\)\no
  &\quad=U(z,y)\(\frac{1}{2}\pd_\m^x-\pd_\m^s+\frac{i}{2}s^\n F_{\m\n}+\frac{i}{12}\(s^\l \pd_\l^x\)s^\n F_{\m\n}\)\(\s^\m \tW(x,s)\),\no
  &\(\tW(z,y)\s^\m U(z,y)\)D_\m^{y\dg}\no
  &\quad=\(\tW(x,s) \s^\m\)\(\frac{1}{2}\overleftarrow{\pd}_\m^x+\overleftarrow{\pd}_\m^s+\frac{i}{2}s^\n F_{\m\n}-\frac{i}{12}\(s^\l \pd_\l^x\)s^\n F_{\m\n}\)U(z,y),
\end{align}
we arrive at the EOM for ${\tW}(x,s)$:
\begin{align}\label{eom_W}
  &\(\frac{1}{2}\pd_\m^x-\pd_\m^s+\frac{i}{2}s^\n F_{\m\n}+\frac{i}{12}\(s^\l \pd_\l^x\)s^\n F_{\m\n}\)\s^\m \tW(x,s)=0, \no
  &\(\frac{1}{2}\pd_\m^x+\pd_\m^s+\frac{i}{2}s^\n F_{\m\n}-\frac{i}{12}\(s^\l \pd_\l^x\)s^\n F_{\m\n}\) \tW(x,s)\s^\m=0.
\end{align}
The kinetic equation is formulated with a quantum distribution function derivable from the Wigner transform of $\tW(x,s)$\cite{Vasak:1987um,Elze:1986qd,Elze:1989un,Zhuang:1995pd}: $W(x,p)=\int \frac{d^4s}{(2\p)^4}e^{-ip\cdot s}\tW(x,s)$, which satisfies the following EOM
\begin{align}\label{eom_wigner}
  &(\frac{1}{2}\D_\m-i\P_\m)\s^\m{W}(x,p)=0, \no
  &(\frac{1}{2}\D_\m+i\P_\m){W}(x,p)\s^\m=0,
\end{align}
where we have defined operators $\D_\m=\pd_\m-\frac{\pd}{\pd p_\n}F_{\m\n}$, $\P_\m= p_\m-\frac{1}{12}\frac{\pd^2}{\pd p_\n\pd p_\l}\frac{\pd}{\pd x^\l}F_{\m\n}$ with the gradient $\frac{\pd}{\pd x^\l}$ acting on $F_{\m\n}$ only. 
We can rewrite \eqref{eom_wigner} into component form by projecting it onto a suitable basis. For right-handed Weyl fermion, $W(x,p)$ is decomposed as
\begin{align}\label{right_decomp}
  W(x,p)=\frac{1}{2}j_\m\bar{\s}^\m.
\end{align}
The projection of \eqref{eom_wigner} gives the following EOM for components
\begin{align}
  \P_{\m} j^{\m}&=0,\label{eom1_o}\\
  \D_{\m} j^{\m}&=0,\label{eom2_o}\\
  \P^{\m} j^{\n} - \P^{\n} j^{\m}&=-\frac{1}{2} \e^{\m\n\r\s} \D_{\r} j_{\s},\label{eom3_o}.
\end{align}
The details of the projection as well as the case for left-handed fermions can be found in appendix A.
By solving the above equations, we can then obtain the current density and stress tensor by momentum integration of $j^\m$:
\begin{align}
  J^\m=&\int d^4p\;\tr\(\s^\m W\)=\int d^4p\,j^\m,\no
  T^{\m\n}=&\frac{1}{2}\int d^4p\;\tr\(p^{\{\m} \s^{\n\}} W\)=\frac{1}{2}\int d^4p\,p^{\{\m} j^{\n\}},
\end{align}
where $X^{\{\m} Y^{\n\}}\equiv X^{\m} Y^{\n}+X^{\n} Y^{\m}$. The contribution of left-handed fermions will be added upon integrating over momenta.

Up to now, we have not specified the order of $F_{\m\n}$ in gradient. We can decompose $F_{\m\n}$ using the fluid velocity as: $F_{\m\n}=\e_{\m\n\r\s}u^\r B^\s+E_\m u_\n-E_\n u_\m$, with $E^\m$ and $B^\m$ being electric and magnetic fields in local rest frame (LRF) of the fluid. For the case of our interest, we consider a strong background magnetic field and a possible electric field perturbation. Thus we regard $B^\m\sim O(\pd^0)$ and $E^\m\sim O(\pd)$. It follows that the $F_{\m\n}$ term in $\P_\m$ can be $O(\pd^0)$. This is the reason to include the second term in $\P_\m$, which counts as $O(\pd^2)$ on the gauge potential.

To proceed, we further choose a constant magnetic field in LRF of the fluid, $B^\m=Bb^\m$ with $B$ and $b^\m$ being magnitude and unit vector both constants in spacetime. While this choice is not the most general situation, it allows us to study the magneto-vortical effect in this simple setting. In particular, it adopts a simple covariant zeroth order solution as \cite{Lin:2019fqo}
\begin{align}\label{cov_sol}
  j_{(0)}^{\m}=&(u+b)^{\m} \d(p\cdot (u+b)) f(p\cdot u) e^{ \frac{p_T^2}{B} }\equiv(u+b)^{\m}j,
\end{align}
where $p_T$ is the momentum component transverse to $u$ and $b$. It is defined by the transverse projector $P^{\m\n}\equiv -g^{\m\n}+u^{\m}u^{\n}-b^{\m}b^{\n}$ as $p_T^\m \equiv -P^{\m\n}p_\n=p^\m-(p\cdot u)u^\m+(p\cdot b)b^\m$. The distribution function involving energy of fermion is
\begin{align}\label{distr}
  f(p\cdot u)=\frac{2}{(2\p)^3}\sum_{r=\pm}\frac{ r \th(rp\cdot u)}{e^{r \(p\cdot u-\m_{{}_R}\)/T}+1},
\end{align}
where $\m_{{}_R}$ is the chemical potential for right-handed Weyl fermions. We take constant chemical potential $\m_{{}_{R/L}}$ and temperature $T$ for simplicity. 

In the next two sections, we will study first order gradient correction to \eqref{cov_sol} induced by constant transverse electric field and vorticity respectively. The resulting current and stress tensor allow us to study (thermal) Hall effect and magneto-vortical effect respectively. The static solution can also be matched with magnetohydrostatics, which is the static limit of magnetohydrodynamics. We will determine several thermodynamic functions through the matching.

\section{Magnetized plasma with a drift}\label{sec_drift}

In this section, we study the magnetized plasma perturbed by a transverse electric field. This would lead to the development a drift velocity like in the case of magnetized plasma consisting of free fermions. We will see drift velocity appear in the solution. In this drift state, we will find the existence of charge and heat flow in the direction of the drift velocity.

We start by turning on a perturbation $a^\m$ at $O(1)$ in gauge potential which gives an $O(\pd)$ electric field $E_{\m}=f_{\m\n}u^\n$ in the transverse direction, i.e., $E_\m b^\m=0$. Here we have isolated the $O(\pd)$ field strength $f_{\m\n}=E_\m u_\n-E_\n u_\m$ from the $O(1)$ part $F_{\m\n}=\e_{\m\n\r\s}u^\r B^\s$. It is sufficient to consider constant $u^\m$, which allows us to drop gradient terms. The EOM are then modified to
\begin{align}
  &p_\m j_{(1)\cD }^\m=0,\label{eom1E} \\
  &D_\m j_{(1)\cD }^\m-f_{\m\n}\frac{\pd}{\pd p_\n}j_{(0)}^\m=0,\label{eom2E} \\
  &p^{[\m} j_{(1)\cD }^{\n]}=-\frac{1}{2}\e^{\m\n\r\s} \(D_\r j_\s^{(1)\cD }-f_{\r\l}\frac{\pd}{\pd p_\l} j_\s^{(0)}\)\label{eom3E},
\end{align}
with $ D_\m\equiv-\frac{\pd}{\pd p_\n}F_{\m\n}=\e_{\m\n\r\s}Bb^\r u^\s\frac{\pd}{\pd p_\n}$ and $X^{[\m}Y^{\n]}\equiv X^\m Y^\n-X^\n Y^\m$. Below we will solve \eqref{eom1E} through \eqref{eom3E} and match the resulting current and stress tensor with MHD.

\subsection{Solution of CKT for a drift state}

We start by putting down an ansatz at $O(\pd)$ for the drift state solution,
\begin{align}\label{anz_drift}
  j_{(1)\cD }^{\m }=(u+b)^{\m }\,p\cdot u_{(1)}\(\frac{\pd }{\pd(p\cdot u)}G_1+G_2\)+u_{(1)}^{\m } G_3,
\end{align}
with $u_{(1)}^{\m}\equiv\frac{1}{2B}\e^{\m \n \r \s } f_{\n \r } b_{\s }$. It is orthogonal to $E_\m,\,b_\m$ and $u_\m$, and is interpreted as drift velocity.
$G_n\propto \d(p\cdot(u+b))e^{ \frac{p_T^2}{B} }$ are undetermined functions depending on momenta $p\cdot u,\,p\cdot b$ and $p_T^2$. We treat $p\cdot u,\,p\cdot b$ and $p_T^2$ as independent in momenta derivatives. More careful discussions can be found in Appendix B where we take momentum in LRF of fluid $q^\m$ as independent variables \footnote{One can equivalently switch between $p^\m$ and $q^\m$ to arrive at the same final solution.}.

One may think there might be a possible change in the leading order distribution $j_{(0)}^\m$ due to the $O(1)$ perturbation $a^\m$ in gauge potential,
\begin{align}\label{cov_sol_perturb}
  j_{(0)\cD }^{\m}=(u+b)^{\m}\(j+a^\l u_\l\,G_4\).
\end{align}
Note that $a^\l u_\l$ can be interpreted as a shift on the chemical potential. Such a contribution is possible, but does not lead to charge/heat current in the direction of the drift velocity. We will not consider this possibility below.
Now we work on the response to the external field $E^\m$ at the first order. The two scalar equations \eqref{eom1E} and \eqref{eom2E} give
\begin{align}
  \eqref{eom1E}\quad\to\quad& p\cdot u_{(1)}p\cdot (u+b)\(\frac{\pd }{\pd(p\cdot u)}G_1+G_2\)+p\cdot u_{(1)} G_3=0\label{eom1_drift}\\
  \eqref{eom2E}\quad\to\quad& -F_{\m\l }\frac{\pd}{\pd p_\l}j_{(1)\cD }^{\m }-f_{\m \n } \frac{\pd }{\pd p_{\n }}j_{(0)}^{\m }\no
  &=B\e_{\m \l \a \b }b^{\a }u^{\b }\frac{\e^{\m \n \r \s } f_{\n \r } b_{\s }}{2B}\frac{2p_T^{\l }}{B}G_3-(u+b)^{\m } f_{\m \n }\frac{2p_T^{\n }}{B}j\no
  &=p^{[\r } u^{\n ]}f_{\n \r } \frac{1}{B}G_3-p^{[\n } u^{\m ]}f_{\m \n } \frac{1}{B}j=0,\label{eom2_drift}
\end{align}
where we have used $E^\m u_\m=0$ and $b^\m f_{\m\n}=0$ in \eqref{eom2_drift}. One finds \eqref{eom1_drift} and \eqref{eom2_drift} are satisfied by $G_1=G_3=j$ and $G_2\propto \d(p\cdot(u+b))$.

We simplify the anti-symmetric tensor equation \eqref{eom3E} as follows.
\begin{align}\label{lhs_E}
  \text{LHS}=&p^{[\m }j_{(1)\cD }^{\n ]}=p\cdot u_{(1)}p^{[\m }(u+b)^{\n ]}\(\frac{\pd }{\pd(p\cdot u)}G_1+G_2\)+p^{[\m } u_{(1)}^{\n ]}G_3\no
  =&p\cdot u_{(1)}\(p_T^{[\m }(u+b)^{\n ]}-p\cdot(u+b)b^{[\m}u^{\n]}\)\(\frac{\pd }{\pd(p\cdot u)}G_1+G_2\)\no
  &\quad+p_T^{[\m } u_{(1)}^{\n ]}G_3+p\cdot u(u+b)^{[\m } u_{(1)}^{\n ]}G_3,\\
  \text{RHS}=&-\frac{1}{2}\e^{\m \n \r \s }\(-F_{\r\l}\frac{\pd}{\pd p_\l} j^{(1)\cD }_{\s }-f_{\r \l }\frac{\pd }{\pd p_{\l }}j_{\s }^{(0)}\).
\end{align}
For two parts on the RHS, the first one can be written as
\begin{align}\label{rhs_E1}
  &-\frac{1}{2}\e^{\m \n \r \s }B\e_{\r \l \a \b }b^{\a }u^{\b }\frac{\pd }{\pd p_{\l }}\[p\cdot u_{(1)}(u+b)_{\s }\(\frac{\pd }{\pd(p\cdot u)}G_1+G_2\)+u^{(1)}_{\s } G_3\]\no
  &=\(p\cdot u_{(1)}p_T^{[\m }(u+b)^{\n ]}+\frac{B}{2}u_{(1)}^{[\m }(u+b)^{\n ]}\)\(\frac{\pd }{\pd(p\cdot u)}G_1+G_2\)+b^{[\m }u^{\n ]}p\cdot u_{(1)}G_3,
\end{align}
where we have used $u\cdot u_{(1)}=b\cdot u_{(1)}=0$. Noting $f_{\r\l}b^\l=0$, the second part writes
\begin{align}\label{rhs_E2}
  &\frac{1}{2}\e^{\m \n \r \s }f_{\r\l}(u+b)_{\s }\(u^{\l }\frac{\pd }{\pd(p\cdot u)}+\frac{2p_T^{\l }}{B}\)j=\frac{B}{2}(u+b)^{[\m }u_{(1)}^{\nu ]}\frac{\pd }{\pd(p\cdot u)}j+p_T^{[\m} u_{(1)}^{\n]}j,
\end{align}
where we have used the following identities shown in Appendix C,
\begin{align}\label{identity}
  \e^{\m \n \r \s } f_{\r \l } (u+b)_{\s } u^{\l }=&B(u+b)^{[\m}u_{(1)}^{\n]},\no
  \e^{\m \n \r \s } f_{\r \l } (u+b)_{\s } p_T^{\l }=&B p_T^{[\m }u_{(1)}^{\n]}.
\end{align}
We collect the LHS and RHS from \eqref{lhs_E}\eqref{rhs_E1}\eqref{rhs_E2} and group them into $b^{[\m}u^{\n]},\,p_T^{[\m}(u+b)^{\n]},\,(u+b)^{[\m}u_{(1)}^{\n]}$ and $p_T^{[\m}u_{(1)}^{\n]}$ terms to fix $G_n$ by comparing the coefficients of the groups. For $b^{[\m}u^{\n]}$ terms, one gets
\begin{align}
  -p\cdot u_{(1)}p\cdot (u+b)\(\frac{\pd }{\pd(p\cdot u)}G_1+G_2\)=p\cdot u_{(1)}G_3,
\end{align}
which holds by $G_1=G_3=j$. The coefficients of $p_T^{[\m}(u+b)^{\n]}$ on two sides cancel out automatically.
For the $(u+b)^{[\m}u_{(1)}^{\n ]}$ terms, we get
\begin{align}
  p\cdot u\,G_3=-\frac{B}{2}\(\frac{\pd }{\pd(p\cdot u)}G_1+G_2\)+\frac{B}{2}\frac{\pd }{\pd(p\cdot u)}j,
\end{align}
which, with $G_1=G_3=j$, gives $G_2=\frac{-2p\cdot u}{B}j$.
The coefficients of $p_T^{[\m } u_{(1)}^{\n ]}$ give $G_3=j$. In summary, the full drift solution for right-handed fermions is
\begin{align}\label{drift_sol}
  j_{(1)\cD }^{\m }=&p\cdot u_{(1)}(u+b)^{\m }\(\frac{\pd }{\pd(p\cdot u)}-\frac{2p\cdot u}{B}\)j+u_{(1)}^\m j.
\end{align}
In fact, up to $O(\pd)$ the solution can be combined with the zeroth order solution into a more suggestive form
\begin{align}\label{drift_sol_compact}
  j_{(0)}^{\m }+j_{(1)\cD }^{\m }=&(u_\cD+b)^{\m }\d(p\cdot(u_\cD+b))f(p\cdot u_\cD)e^{(p^2-(p\cdot u_\cD)^2+(p\cdot b)^2)/B}.
\end{align}
This is nothing but the zeroth order solution with $u^\m\to u_\cD^\m\equiv(u+u_{(1)})^\m$.
The counterpart for left-handed fermions can be obtained by sending $b\to-b$ and $\m_{{}_R}\to\m_{{}_L}$.

We may either choose $u_\cD^\m$ or $u^\m$ as fluid velocity, which correspond to different frame choices in hydrodynamics. In the former case the in medium electric field defined by $u_\cD$ is vanishing $f_{\m\n}u^\n+F_{\m\n}u_{(1)}^\n=0$. It follows that there is no charge/heat current orthogonal to the fluid velocity. This corresponds to the Landau frame. The latter case contains both charge/heat current. As we will see below, it can be matched with the constitutive equations of MHD in thermodynamic frame \cite{Hernandez:2017mch}.

\subsection{Matching with magnetohydrodynamics}

With \eqref{cov_sol} and \eqref{drift_sol}, we are ready to calculate the current and stress tensor by momenta integration. Here we simply collect the final results and leave the details of the evaluation to appendix B.
\begin{align}
  J_{(0)}^\m=&\frac{\m B}{2\p^2}u^\m+\frac{\m_5 B}{2\p^2}b^\m,\label{J0}\\
  T_{(0)}^{\m\n}=&\frac{\c_{{}_V}B}{2\p^2}(u^\m u^\n+b^\m b^\n)+\frac{\c_{{}_A}}{2\p^2}u^{\{\m }b^{\n\}},\label{T0}\\
  J_{(1)\cD }^\m=&-\frac{\m }{2\p^2} \e^{\m \n \r \s } u_{\n } E_{\r } b_{\s },\label{J1_d}\\
  T_{(1)\cD }^{\m\n}=&-\frac{1}{2\p^2}\(\frac{B}{4}+\chi \)u^{\{\m}\e^{\n\} \l \r \s }u_{\l } E_{\r } b_{\s },\label{T1_d}
\end{align}
where we have defined $\c_{{}_V}\equiv\frac{\m^2+\m_5^2}{2}+\frac{\p^2 T^2}{6}$ and $\c_{{}_A}\equiv\m \m_5$.
We see that \eqref{J0} contains charge density and current density contributions. From the charge density, we easily recognize the charge susceptibility $\c_\m=\frac{B}{2\p^2}$, which is given by density of LLL states. The current density is the celebrated CME result. \eqref{T0} is effectively reduced to $1+1$ dimensional in the LLL approximation and there is net longitudinal heat flow in the presence of $\m_5$. As we stressed the in medium electric field $E_\m=F_{\m\n}u^\n$ in thermodynamic frame leads to Hall current and heat current in \eqref{J1_d} and \eqref{T1_d}.

To match with constitutive equations of MHD, which relates components of current and stress tensor through thermodynamic functions, we closely follow the notations of \cite{Hernandez:2017mch}, in which the current and stress tensor are decomposed as \footnote{In making the comparison, we note that \cite{Hernandez:2017mch} uses a different signature in metric. Also their definition of electromagnetic field or alternatively current differs from ours by a sign. We quote the converted constitutive relations of MHD.}
\begin{align}
  J^\m=&\cN u^\m+\cJ^\m\\
  T^{\m\n}=&\cE u^\m u^\n+\cP\D^{\m\n}+\cQ^\m u^\n+\cQ^\n u^\m+\cT^{\m\n}
\end{align}
where $\D^{\m\n}\equiv -g^{\m\n}+u^\m u^\n=P^{\m\n}+b^\m b^\n$. One has $\cN= u_\m J^\m$, $\cJ_\m= -\D_{\m\l}J^\l$, $\cE= u_\m u_\n T^{\m\n}$, $\cP= \frac{1}{3}\D_{\m\n}T^{\m\n}$, $\cQ_\m= -\D_{\m\a}u_\b T^{\a\b}$ and $\cT^{\m\n}= \frac{1}{2}\(\D_{\m\a}\D_{\n\b}+\D_{\n\a}\D_{\m\b}-\frac{2}{3}\D_{\m\n}\D_{\a\b}\)T^{\a\b}$. From \eqref{J0} through \eqref{T1_d}, we obtain the components of the current and stress tensor at $O(1)$,
\begin{align}\label{vor_sol_0}
  &\cN_{(0)}=\frac{\m B}{2\p^2},\qquad\qquad
  \cJ_{(0)}^\m=\frac{\m_5 B}{2\p^2}b^\m,\no
  &\cE_{(0)}=\frac{\c_{{}_V} B}{2\p^2},\qquad\qquad
  \cQ_{(0)}^\m=\frac{\c_{{}_A} B}{2\p^2}b^\m,\no
  &\cP_{(0)}=\frac{\c_{{}_V} B}{6\p^2},\qquad\qquad
  \cT_{(0)}^{\m\n}=\frac{\c_{{}_V} B}{6\p^2}\(2b^\m b^\n-P^{\m \n}\),
\end{align}
and two nonvanishing parity odd components at $O(\pd)$,
\begin{align}\label{heat_drift}
  \cJ_{(1)\cD }^{\m }=&-\frac{\m }{2\p^2} \e^{\m \n \r \s } u_{\n } E_{\r } b_{\s },\no
  \cQ_{(1)\cD }^\m=&-\frac{1}{2\p^2}\(\frac{B}{4}+\chi \)\e^{\m \n \r \s }u_{\n } E_{\r } b_{\s }.
\end{align}
The spatial current $\cJ_{(1)\cD }^{\m }$ along the drift velocity gives the Hall conductivity $\s_H=\frac{\m}{2\p^2}$. The heat flow $\cQ_{(1)\cD }^\m$ is parallel to the Hall current. They are nonvanishing in the absence of $\m_5$.

Meanwhile, the constitutive relations for components of MHD \cite{Hernandez:2017mch} give,
\begin{align}\label{cons_rela_d}
  \cN_{(0)}=&n=p_{,\m}\,,\qquad\qquad\qquad\qquad
  \cP_{(0)}=\P=p-\frac{4}{3}p_{,B^2}B^2,\no
  \cE_{(0)}=&\e=-p+Tp_{,{}_T}+\m p_{,\m}\,,\qquad
  \cT_{(0)}^{\m\n}=\frac{1}{3}\a_{{}_{BB}}B^2\(2b^\m b^\n-P^{\m \n}\),
\end{align}
and
\begin{align}\label{heat_mhd}
  \cJ_{(1)}^\m=&-\a_{{}_{BB},\m}\e^{\m \n \r \s }u_{\n } E_{\r } B_{\s },\no
  \cQ_{(1)}^\m=&\(M_{\o ,\m }+2p_{,B^2}\)\e^{\m \n \r \s }u_{\n } E_{\r } B_{\s },
\end{align}
where $p$ is pressure, $\a_{{}_{BB}}=2 p_{,B^2}$ is magnetic susceptibility and $M_\o$ is magneto-vortical susceptibility. Note that $p$ and $\P$ are thermodynamic functions here, not to be confused with indexed $p_\m$ and $\P_\m$.
To compare with MHD, we mute $\m_5$ to get $\c_{{}_V}\to\c\equiv\frac{\m^2}{2}+\frac{\p^2 T^2}{6}$ and vanishing parity odd coefficient $\c_{{}_A}$. Then one easily finds the $O(1)$ components in \eqref{vor_sol_0} satisfy the constitutive relations in \eqref{cons_rela_d} by taking $p=\c B/2\p^2$. At $O(\pd)$, by matching \eqref{heat_drift} and \eqref{heat_mhd}, we can fix $M_\o$ in drift state as
\begin{align}\label{MomegaD}
  M_{\o }^{\cD}=-\frac{\m}{8\p^2}-\frac{\x}{2\p^2B}.
\end{align}

\section{Magnetized plasma with a vorticity}\label{sec_vortical}

In this section, we study the effects of a steady vorticity parallel to the magnetic field in the plasma. We turn on a vorticity $\o^\m=\frac{1}{2}\e^{\m\n\r\s}u_\n\pd_\r u_\s=\o b^\m$ in the fluid along the direction of the magnetic field with $\o=-\o^\m b_\m$. We further require the absence of shear or bulk tensors in the fluid. Then we solve \eqref{eom1_o}-\eqref{eom3_o} to the first order of gradient, or equivalently, $O(\o)$. The solution is to be referred to as vortical solution. In matching the resulting current and stress tensor with MHD, we find one of the thermodynamic functions $M_\o$ has a different value from \eqref{MomegaD}. The apparent discrepancy will be resolved with a reinterpretation of the results, which precisely corresponds to shift of chemical potential discussed in the introduction.

\subsection{Vortical Solution}

Denoting the first order solution by $j_{(1)\cV }^\m$, we can write the equations explicitly as
\begin{align}
  &p_\m j_{(1)\cV }^\m+\d\P_\m j_{(0)}^\m=0,\label{eom1} \\
  &\pd_\m j_{(0)}^\m+ D_\m j_{(1)\cV }^\m=0,\label{eom2} \\
  &p^{[\m} j_{(1)\cV }^{\n]}+\d\P^{[\m} j_{(0)}^{\n]}=-\frac{1}{2}\e^{\m\n\r\s}\(\pd_\r j_\s^{(0)}+ D_\r j_\s^{(1)\cV}\)\label{eom3}.
\end{align}
where we have defined $\d\P_\m\equiv \P_\m-p_\m$. Here we choose $ D_\m=\e_{\m\n\r\s}Bb^\r u^\s\frac{\pd}{\pd p_\n}$ corresponding to a constant magnetic field in the LRF of the fluid. The field strength $F_{\m\n}=\e_{\m\n\r\s}Bb^\r u^\s$ is spacetime dependent through the fluid velocity. In \eqref{eom1}-\eqref{eom3}, $j_{(1)\cV }^\m$ is sourced by terms proportional to $\pd\,j_{(0)}$ and $\d\P\,j_{(0)}$. The former captures the spacetime derivatives on the distribution function and the latter is the counterpart in field strength. While mathematically they both reduce to spacetime dependence of fluid velocity, their physical difference is clear. Accordingly we will split $j_{(1)\cV }^\m$ into two parts
\begin{align}
  j_{(1)\cV }^\m=j_{(1)\cC }^\m+j_{(1)\cA }^\m,
\end{align}
with $j_{(1)\cC }^\m$ and $j_{(1)\cA }^\m$ satisfying the following EOM
\begin{align}
  &p_\m j_{(1)\cC }^\m=0,\label{eom1C} \\
  &\pd_\m j_{(0)}^\m+ D_\m j_{(1)\cC }^\m=0,\label{eom2C} \\
  &p^{[\m} j_{(1)\cC }^{\n]}=-\frac{1}{2}\e^{\m\n\r\s}\(\pd_\r j_\s^{(0)}+ D_\r j_\s^{(1)\cC }\),\label{eom3C}\\
  &p_\m j_{(1)\cA }^\m+\d\P_\m j_{(0)}^\m=0,\label{eom1F} \\
  & D_\m j_{(1)\cA }^\m=0,\label{eom2F} \\
  &p^{[\m} j_{(1)\cA }^{\n]}+\d\P^{[\m} j_{(0)}^{\n]}=-\frac{1}{2}\e^{\m\n\r\s}\( D_\r j_\s^{(1)\cA }\),\label{eom3F}
\end{align}
Similar to \eqref{anz_drift}, we take the following ansatz for $j_{(1)\cC }^\m$,
\begin{align}\label{anz_vortical_C}
  j_{(1)\cC }^{\m }=&(u+b)^{\m }\frac{\o\,p\cdot p_T }{B}\(\frac{\pd }{\pd(p\cdot u)}F_1+ F_2\)+\frac{\o\,p_T^{\m } }{B} F_3+(u+b)^{\m }F_4,
\end{align}
where $F_n\propto \d(p\cdot(u+b))e^{ \frac{p_T^2}{B} }$ are undetermined functions.
Noting the on shell condition $\d(p\cdot(u+b))$ in $F_n$, \eqref{eom1C} gives
\begin{align}
  p_{\m } j_{(1)\cC }^{\m }=&p\cdot(u+b)\(\frac{\o\,p_T^2 }{B}\(\frac{\pd }{\pd(p\cdot u)}F_1+ F_2\)+F_4\)+p_\m \frac{\o\,p_T^\m}{B} F_3\no
  =&\frac{\o }{B}\(-p_T^2 F_1+p_T^2 F_3\)=0,\no
  \to\quad&-F_1+F_3=0,
\end{align}
where we have used integration by parts for the $\frac{\pd }{\pd(p\cdot u)}F_1$ term and $p_{\m }p_T^{\m }=p_T^2$.

We start with \eqref{eom2C}, which can be simplified using the bulk free condition $\pd_\m u^\m=0$. In this case, \eqref{eom2C} becomes
\begin{align}
  &\pd_\m j_{(0)}^{\m }- D_\m j_{(1)\cC }^{\m }=(u+b)^{\m }p^{\l }\pd_\m u_{\l }\(\frac{\pd }{\pd(p\cdot u)}-\frac{2p\cdot u}{B}\)j\no
  &\quad+B \e_{\m \n \r \s }b^{\r }u^{\s }\frac{\pd}{\pd p_\n}\[(u+b)^{\m }\(\frac{\o\,p_T^2 }{B}\(\frac{\pd }{\pd(p\cdot u)}F_1+ F_2\)+F_4\)+\frac{\o\,p_T^\m}{B} F_3\]=0.
\end{align}
Furthermore, with $\pd_\r u_{\s }=-\o b^{\m } u^{\n } \e_{\m \n \r \s }$ following from the shear free condition, one finds all the terms vanish by anti-symmetry of $\e_{\m \n \r \s }$. Therefore \eqref{eom2C} is automatically satisfied.

The anti-symmetric tensor equation requires some work. Firstly, we simplify the left hand side (LHS) and right hand side(RHS) of \eqref{eom3C} as follows. The LHS writes
\begin{align}\label{lhs_C}
  &p^{[\m} j_{(1)\cC }^{\n]}=p^{[\m }(u+b)^{\n ]}\(\frac{\o\,p_T^2 }{B}\(\frac{\pd }{\pd(p\cdot u)}F_1+ F_2\)+F_4\)+\frac{\o }{B}p^{[\m }p_T^{\n ]} F_3\no
  &=\(p_T^{[\m }(u+b)^{\n ]}-p\cdot (u+b)b^{[\m }u^{\n ]}\)\(\frac{\o\,p_T^2 }{B}\(\frac{\pd }{\pd(p\cdot u)}F_1+ F_2\)+F_4\)\no
  &\qquad-\frac{\o }{B}p\cdot u\, p_T^{[\m }(u+b)^{\n ]}F_1,
\end{align}
where we have used $p_T^{[\m }(u+b)^{\n ]}=p^{[\m }(u+b)^{\n ]}+p\cdot (u+b)b^{[\m }u^{\n ]}$ in the first term and $p^{[\m }p_T^{\n ]}=-p\cdot u\, p_T^{[\m }(u+b)^{\n ]}$ by on shell condition in the second term. \eqref{lhs_C} contains two independent structures $p_T^{[\m }(u+b)^{\n ]}$ and $b^{[\m }u^{\n ]}$, which are transverse-longitudinal and longitudinal-temporal types.
There are two parts on the RHS,
\begin{align}
  -\frac{1}{2}\e^{\m\n\r\s}\pd_\r j_\s^{(0)}-\frac{1}{2}\e^{\m\n\r\s}D_\r j_\s^{(1)\cC }.
\end{align}
Using the relation $\e^{\m \n \r \s } \pd_\r u_{\s }=2 \o \(b^{\m }u^{\n }-b^{\n }u^{\m }\)$ and $\pd_\r u_{\l }=-\o \e_{\r \l \a \b }b^{\a }u^{\b }$, we can simplify the first term as
\begin{align}\label{rhs_C1}
  &-\frac{1}{2}\e^{\m \n \r \s } \[\pd_\r u_{\s } +(u+b)_{\s }p^{\l }\pd_\r u_{\l }\(\frac{\pd }{\pd(p\cdot u)}-\frac{2p\cdot u}{B}\)\]j\no
  &=-\o \[b^{[\m } u^{\n ]}-\frac{1}{2}\e^{\m \n \r \s }(u+b)_{\s }p^{\l }\e_{\r \l \a \b }b^{\a }u^{\b }\(\frac{\pd }{\pd(p\cdot u)}-\frac{2p\cdot u}{B}\)\]j\no
  &=-\o\[b^{[\m } u^{\n ]}j+\frac{1}{2}p_T^{[\m } (u+b)^{\n ]}\(\frac{\pd j}{\pd(p\cdot u)}-\frac{2p\cdot u}{B}j\)\].
\end{align}
In the second part, given that $F_n\propto e^{ \frac{p_T^2}{B} }$ depend on momenta by $p\cdot u$, $p\cdot b$ and $p_T^2$ only, we note when acting on $j_{\s }^{(1)\cC }$, the operator $\frac{\pd }{\pd p_{\l }}$ can pull out terms like $u^\l\frac{\pd}{\pd(p\cdot u)}$, $b^\l\frac{\pd}{\pd p\cdot b}$, $p_T^\l\frac{\pd}{\pd p_T^2}$ and $\d_\s^\l$, where only the last two cases survive upon contraction with $\e_{\r\l\a\b }$. One gets
\begin{align}\label{rhs_C2}
  &-\frac{1}{2}\e^{\m \n \r \s }\e_{\r \l \a \b }B b^{\a }u^{\b }\frac{\pd }{\pd p_{\l }}\[(u+b)_{\s }\(\frac{\o\,p_T^2 }{B}\(\frac{\pd }{\pd(p\cdot u)}F_1+ F_2\)+F_4\)+\frac{\o\,p^T_{\s }}{B} F_3\]\no
  &=-\frac{1}{2}\e^{\m \n \r \s }\e_{\r \l \a \b }B b^{\a }u^{\b }\bigg[(u+b)_{\s }\frac{2p_T^\l}{B}\(\o\,\(1+\frac{p_T^2}{B}\)\(\frac{\pd }{\pd(p\cdot u)}F_1+F_2\)+F_4\)\no
  &\quad+\frac{\o}{B}\(\d_\s^\l+\frac{2p_T^\l p_\s^T}{B}\) F_1\bigg]\no
  &=p_T^{[\m } (u+b)^{\n ]}\(\o\(1+\frac{p_T^2}{B}\)\(\frac{\pd }{\pd(p\cdot u)}F_1+F_2\)+F_4\)+b^{[\m } u^{\n ]}\o\(1+\frac{p_T^2}{B}\)F_1.
\end{align}
The RHS from \eqref{rhs_C1}\eqref{rhs_C2} contains the same structures as the LHS. By matching the coefficient of $b^{[\m }u^{\n ]}$ using integration by part, we can fix $F_1=j$.
The remaining structure reads
\begin{align}\label{lhs_rhs_C}
  &\text{LHS$-$RHS}
  =-\o\,p_T^{[\m } (u+b)^{\n ]}\bigg[\frac{2p\cdot u }{B}j+\(\frac{1}{2}\frac{\pd }{\pd(p\cdot u)}j+F_2\)\bigg].
\end{align}
We note that $F_4$ cancels in \eqref{lhs_rhs_C}. In fact, $F_4\propto\d(p\cdot(u+b))e^{ \frac{p_T^2}{B} }$, which can be recognized as the change of distribution function.
We also note that $F_2\propto \d(p\cdot(u+b))$ while $\frac{\pd}{\pd(p\cdot u)}j$ contains $\d'(p\cdot(u+b))$, which immediately shows \eqref{lhs_rhs_C} cannot be identically zero. This will be resolved only after we combine with the solution $j_{(1)\cA }^\m$.

To solve for $j_{(1)\cA }^\m$, we note that \eqref{eom1F}-\eqref{eom3F} can be formally obtained from the zeroth order by the replacement $p_\m\to p_\m+\d\P_\m$ and $j_{(0)}^\m\to j_{(0)}^\m+j_{(1)\cA }^\m$ and expanded to $O(\pd)$. The formal solution motivates the following ansatz 
\begin{align}\label{anz_vortical_F}
j_{(1)\cA }^{\mu }=(u+b)^{\m} \d'\(p\cdot (u+b)\)(u+b)\cdot\d\P\,\tj,
\end{align}
where $\tj\equiv f(p\cdot u)\,e^{ \frac{p_T^2}{B} }$. $\d\P$ is a differential operator, whose explicit expression is worked out in appendix B as
\begin{align}\label{Pis}
  \d\P_\m&=u_\m \d\P_u+\frac{2 p_\m^T}{B} \d\P_T\quad\text{with}&\no
  \d\P_u&=\frac{B\o}{12}P_{\l\n}\frac{\pd }{\pd p_\l}\frac{\pd }{\pd p_\n},\quad
  \d\P_T=\frac{B\o}{12}\frac{\pd }{\pd (p\cdot u)},
\end{align}
up to $O(\o)$.
Below we verify \eqref{anz_vortical_F} gives an extra contribution that cancels out the $\frac{\pd }{\pd(p\cdot u)}j$ term in \eqref{lhs_rhs_C} and fixes $F_2$ to gives a proper final solution at $O(\o)$. Using $p_T\cdot u=p_T\cdot b=0$, \eqref{eom1F} gives
\begin{align}
  &p_\m j_{(1)\cA }^\m+\(u_\m \tP+\frac{2 p_\m^T}{B}\bP\)j_{(0)}^{\mu }\no
  &=p\cdot (u+b) \d'\(p\cdot (u+b)\)\d\P_u\,\tj+\d\P_u\(\d\(p\cdot (u+b)\)\,\tj\)=0.
\end{align}
Note that $\d\P_u$ involves differentiation on the transverse momenta $p_T^\m$ and therefore does not act on $p\cdot u,\,p\cdot b$, which means $\d\P_u \d\(p\cdot (u+b)\)=\d\(p\cdot (u+b)\)\d\P_u $. We can then see the above equation holds upon integration by parts. 
By the anti-symmetric $\e_{\m\n\r\s}b^\r u^\s$ term in $ D_\m$, \eqref{eom2F} is trivially satisfied as $j_{(1)\cA }^\m\propto (u+b)^\m$.
We then proceed to the  anti-symmetric tensor equation \eqref{eom3F} as follows. Explicitly, the LHS of \eqref{eom3F} writes
\begin{align}\label{lhs_F}
  \d\P^{[\m} j_{(0)}^{\n]} +p^{[\m}j_{(1)\cA }^{\n]}=&\(u^{[\m} b^{\n]} \d\P_u+\frac{2}{B} p_T^{[\m}(u+b)^{\n]}\d\P_T\)\(\d\(p\cdot (u+b)\) \,\tj\)\no
  &\quad+p^{\m} (u+b)^{\n} \(\d'\(p\cdot (u+b)\) \d\P_u\,\tj\)\no
  =&\(\frac{2}{B} p_T^{[\m}(u+b)^{\n]}\d\P_T-b^{[\m} u^{\n]} \d\P_u\)\(\d\(p\cdot (u+b)\) \,\tj\)\no
  &\quad+\(p_T^{\m} (u+b)^{\n}-p\cdot(u+b)b^{[\m} u^{\n]}\) \(\d'\(p\cdot (u+b)\) \d\P_u\,\tj\)\no
  =&p_T^{[\m}(u+b)^{\n]}\(\frac{2}{B}\d\P_T\(\d\(p\cdot (u+b)\) \,\tj\)+\d'\(p\cdot (u+b)\) \d\P_u\,\tj\),
\end{align}
where we have canceled out the $b^{[\m} u^{\n]}$ terms using integration by parts in the last equality. The RHS of \eqref{eom3F} gives 
\begin{align}\label{rhs_F}
  -\frac{1}{2}\e^{\m\n\r\s}\frac{\pd }{\pd p_{\l }}\e_{\r \l \a\b}Bb^{\a}u^{\b}j_{\s}^{(1)\cA }=&-\frac{1}{2}\e^{\m\n\r\s}\e_{\r \l \a\b}Bb^{\a}u^{\b}(u+b)_{\s }\d'\(p\cdot(u+b)\) \d\P_u \frac{\pd }{\pd p_{\l }} \tj\no
  =&p_T^{\m}(u+b)^{\n} \d'\(p\cdot(u+b)\)\(\d\P_u-\frac{\o}{3}\)\tj,
\end{align}
where the $\frac{\o}{3}$ term in the last equality comes from the commutator $[\d\P_u,p_T^\l]$. Now, gathering \eqref{lhs_F} and \eqref{rhs_F}, we have 
\begin{align}\label{lhs_rhs_F}
  \text{LHS$-$RHS}=&p_T^{[\m}(u+b)^{\n]}\(\frac{2}{B}\d\P_T\(\d\(p\cdot (u+b)\) \,\tj\)+\frac{\o}{3}\d'\(p\cdot (u+b)\)\,\tj\)\no
  =&p_T^{[\m}(u+b)^{\n]}\(\frac{\o}{6}\d\(p\cdot (u+b)\)f'(p\cdot u)e^{\frac{p_T^2}{B}}+\frac{\o}{2}\d'\(p\cdot (u+b)\)\,\tj\).
\end{align}
Cancellation of \eqref{lhs_rhs_C} and \eqref{lhs_rhs_F} requires
\begin{align}
  F_2=-\frac{1}{3} \d\(p\cdot (u+b)\) f'(p\cdot u)e^{\frac{p_T^2}{B}}-\frac{2p\cdot u}{B}j.
\end{align}
Combining \eqref{anz_vortical_C} and \eqref{anz_vortical_F}, we have the following solution up to possible addition of $F_4$ as
\begin{align}\label{vortical_F2}
  &j_{(1)\cV }^{\m }=(u+b)^{\m}\bigg[ -\frac{\o}{3}\(\frac{p_T^2}{B}+1\)\d'\(p\cdot (u+b)\)f(p\cdot u)+\frac{2\o p_T^2}{3B}\d\(p\cdot (u+b)\)f'(p\cdot u) \no
  &\qquad\qquad-\frac{2\o p_T^2}{B^2}p\cdot u\,\d\(p\cdot (u+b)\)f(p\cdot u)\bigg]e^{\frac{p_T^2}{B}}+\frac{\o p_T^\m}{B}\d\(p\cdot (u+b)\)f(p\cdot u)e^{\frac{p_T^2}{B}}.
\end{align}
The above procedure can be easily generalized to the case of left-handed fermions with the solution given by the replacement $b\to-b$ and $\m_{{}_R}\to\m_{{}_L}$\footnote{Here it is more appropriate to regard $b$ as the spin direction of LLL states rather than the magnetic field direction.}.

\subsection{Matching with magnetohydrodynamics}

Again, after integration over momenta and  summation over right/left-handed contributions detailed in Appendix B, \eqref{vortical_F2} gives the current and stress tensor as
\begin{align}
  J_{(1)\cV }^\m=&\frac{\o}{2\p^2}\(2\c_{{}_V}+\frac{2}{3}B\)u^\m+\frac{\o}{2\p^2}2\c_{{}_A} b^\m,\label{J1_v}\\
  T_{(1)\cV }^{\m\n}=&\frac{\o}{2\p^2}\(2\x_{{}_V}+\frac{2}{3}\m B\)u^\m u^\n+\frac{\o}{2\p^2}\(2\x_{{}_V}-\frac{1}{3}\m B\)b^\m b^\n\no
  &+\frac{\o}{2\p^2}\(2\x_{{}_A}+\frac{1}{6}\m_5B\)u^{\{\m} b^{\n\}}+\frac{\o}{2\p^2}\frac{\m B}{2}P^{\m\n},\label{T1_v}
\end{align}
where we have defined $\x_{{}_V}\equiv\frac{1}{3} \m \(\m^2+3\m_5^2+\p^2 T^2\)$ and $\x_{{}_A}\equiv\frac{1}{3} \m_5 \(\m_5^2+3\m^2+\p^2 T^2\)$. 
We note that the current density in \eqref{J1_v} is in agreement with CVE. The charge density does contains an $O(B^0)$ medium contribution and $O(B)$ vacuum contribution. The latter however contradicts \eqref{mve}. The contradiction should not be a surprise. The reason is our vortical solution \eqref{vortical_F2} is unique only up to possible addition of $F_4$, which we have not considered so far.
In fact, since we consider the magnetized plasma in a steady vorticity, the state is not reached as a response to vorticity, thus we do not have a first principle to fix $F_4$ within our approach.
We can choose any $F_4\propto \d(p\cdot(u+b))e^{\frac{p_T^2}{B}}g(p\cdot u)$, which necessarily modifies $J_{(1)\cV }^\m$ and $T_{(1)\cV }^{\m\n}$.

Fortunately the ambiguity can still be fixed by matching with constitutive equations of MHD. From \eqref{J1_v} and \eqref{T1_v}, we obtain the components of the current and stress tensor at $O(\pd)$,
\begin{align}\label{vor_sol_1}
  &\cN_{(1)\cV }=\frac{\o}{2\p^2}\(2\c_{{}_V}+\frac{2}{3}B\),\qquad
  \cJ_{(1)\cV }^\m=\frac{\o}{2\p^2}2\c_{{}_A}b^\m,\no
  &\cE_{(1)\cV }=\frac{\o}{2\p^2}\(2\x_{{}_V}+\frac{2}{3}\m B\),\qquad
  \cQ_{(1)\cV }^\m=\frac{\o}{2\p^2}\(2\x_{{}_A}+\frac{1}{6}\m_5 B\)b^\m,\no
  &\cP_{(1)\cV }=\frac{\o}{6\p^2}\(2\x_{{}_V}+\frac{2}{3}\m B\),\qquad
  \cT_{(1)\cV }^{\m\n}=\frac{\o}{6\p^2}\(2\x_{{}_V}-\frac{5}{6}\m B\)\(2b^\m b^\n-P^{\m \n}\),
\end{align}
for which, the constitutive relations in MHD \cite{Hernandez:2017mch} are
\begin{align}\label{cons_sol_1}
  &\cN_{(1)}=f_\cN=-2\(2p_{,B^2}+M_{\o,\m}\)B\o,\no
  &\cE_{(1)}=f_\cE=-2\(TM_{\o,T}+\m M_{\o,\m}-2M_{\o}\)B\o,\no
  &\cP_{(1)}=f_\cP=\frac{2}{3}\(M_{\o}+4M_{\o,B^2}B^2\)B\o,\no
  &\cT_{(1)}=f_\cT=-\frac{4}{3}\(M_{\o,B^2}B^2+M_\o\)B\o,
\end{align}
with $\cT_{(1)}^{\m\n}=\(2b^\m b^\n-P^{\m \n}\)\cT_{(1)}$. 
To proceed, we turn off $\m_5$ to get $\c_{{}_V}\to\c\equiv\frac{\m^2}{2}+\frac{\p^2 T^2}{6}$, $\x_{{}_V}\to\x\equiv\frac{1}{3} \m \(\m^2+\p^2 T^2\)$ and vanishing parity odd coefficients $\c_{{}_A},\,\x_{{}_A}$. Then the counterparts in CKT are reduced to
\begin{align}\label{cons_sol}
  &\cN_{(1)\cV }=f_\cN'=\frac{\o}{2\p^2}\(2\c+\frac{2}{3}B\),\no
  &\cE_{(1)\cV }=f_\cE'=\frac{\o}{2\p^2}\(2\x+\frac{2}{3}\m B\),\no
  &\cP_{(1)\cV }=f_\cP'=\frac{\o}{6\p^2}\(2\x+\frac{2}{3}\m B\),\no
  &\cT_{(1)\cV }={f'}_\cT=\frac{\o}{6\p^2}\(2\x-\frac{5}{6}\m B\),
\end{align}
up to possible addition of $F_4$, which corresponds to $O(\o)$ modification of distribution. The simplest possible modification is through $O(\o)$ modification of temperature and chemical potential. If this were the case, the effect of $F_4$ can be realized by a frame transformation, which amounts to a redefinition of temperature and chemical potential \cite{Kovtun:2012rj}. We will see below a frame transformation indeed allows for matching with MHD. The matching is most easily done through the following frame invariant variables \cite{Hernandez:2017mch},
\begin{align}\label{frame_inv}
  f\equiv &f_\cP-\(\frac{\pd \P}{\pd \e}\)_n f_\cE-\(\frac{\pd \P}{\pd n}\)_{\e} f_\cN,\no
  t\equiv &f_\cT-\frac{B^2}{3}\[\(\frac{\pd \a_{{}_{BB}}}{\pd \e}\)_n f_\cE+\(\frac{\pd \a_{{}_{BB}}}{\pd n}\)_{\e} f_\cN\].
\end{align}
We should match $f$ and $t$ constructed using \eqref{cons_sol_1} and \eqref{cons_sol} to fix $M_\o$. Using
\begin{align}
  &\e=\frac{\c B}{2\p^2},\quad\P=\frac{\c B}{6\p^2},\quad n=\frac{\m B}{2\p^2},\quad\a_{{}_{BB}}=\frac{\c}{2\p^2B}\label{thermal}\\
  &\quad\Rightarrow\quad\(\frac{\pd \P}{\pd n}\)_\e=0,\quad\(\frac{\pd \P}{\pd \e}\)_n=\frac{1}{3},\quad\(\frac{\pd \a_{{}_{BB}}}{\pd n}\)_\e=0,\quad\(\frac{\pd \a_{{}_{BB}}}{\pd \e}\)_n=\frac{1}{B^2},
\end{align}
in \eqref{frame_inv}, one gets
\begin{align}\label{mat_eqn}
  &\frac{2}{3}\(M_{\o }+4M_{\o ,B^2}B^2\)+\frac{2}{3}\(T M_{\o ,T}+\m M_{\o,\m }-2 M_{\o }\)=0,\no
  &\frac{4}{3}\(M_{\o,B^2}B^2+M_\o\)-\frac{2}{3}\(T M_{\o ,T}+\m M_{\o,\m }-2 M_{\o }\)=\frac{\m}{4\p^2},
\end{align}
which are satisfied by $M_{\o }^{\cV}=\frac{\m}{8\p^2}+\frac{\#\x}{B}$. An arbitrary coefficient $\#$ is allowed in the medium part. By matching the medium part with \eqref{MomegaD}, we fix
\begin{align}\label{MomegaV}
  M_{\o }^{\cV}=\frac{\m}{8\p^2}-\frac{\x}{2\p^2B}.
\end{align}
We see the matching equations \eqref{mat_eqn} are over-determined. The agreement on the medium part of $M_\o$ between drift and vortical solutions is rather non-trivial. The disagreement on the vacuum part needs further clarification.

\subsection{Vacuum ambiguity}
Recall $M_\o$ is defined by the change of free energy in response to magneto-vortical source \cite{Kovtun:2012rj}, which reads in our case
\begin{align}
  \D \cF=-2M_\o B\o.
\end{align}
The definition implicitly assumes the vacuum is not changed as the magneto-vortical source is turned on adiabatically. If the vacuum state is changed in the process, we should instead use the new vacuum state as reference point in calculating the free energy. To reconcile $M_\o^\cD$ and $M_\o^\cV$, the vacuum energy density needs to be lowered by $\frac{\m B\o}{2\p^2}$ in the adiabatic process above. Indeed this is consistent with the picture that each LLL state has a lowered energy \eqref{E-shift}. The chemical potential measured with respect to the lowered vacuum is shifted up by $\D\m=\o$ for particles. $\frac{\m B\o}{2\p^2}$ is accounted by the product of $\D\m$ and charge density $\frac{\m B}{2\p^2}$ of the LLL states.

The new vacuum is given by $j^\m=j_{(0)}^\m+j_{{}_\text{vac}}^\m$, with the shift part reads
\begin{align}\label{j_vac}
  j_{{}_\text{vac}}^{\m}=\o(u\pm b)^\m\d\(p\cdot (u\pm b)\) f'\(p\cdot u\) e^{\frac{p_T^2}{B}},
\end{align}
with upper/lower signs for right/left handed fermions respectively. \eqref{j_vac} has the simple interpretation as from a shift in chemical potential $\d \m_{{}_\text{vac}}=-\o$. The corresponding shift in stress tensor and current are evaluated as
\begin{align}\label{TJ_shift}
  T_{{}_\text{vac}}^{\m\n}=\frac{1}{2}\int\,d^4p p^{\{\m}j_{{}_\text{vac}}^{\n\}}=-\frac{\m B \o }{2 \pi ^2}\(u^\m u^\n+b^\m b^\n\),\quad J_{{}_\text{vac}}^\m=\int\,d^4p j_{{}_\text{vac}}^\m=-\frac{B \o }{2 \p^2}u^\m.
\end{align}
Note that apart from the needed shift in energy density, there is also a negative shift of charge density in the new vacuum.

Now we can calculate the change of charge density using $M_\o^\cV$ and the vacuum shifted density \eqref{TJ_shift} as
\begin{align}\label{final_J}
  \D J^0&=-2(2p_{,B^2}+M_{\o,\m}^\cV)B\o-J_{\text{vac}}^0 =\(\frac{B}{4\p^2}+\frac{\c}{\p^2}\)\bo\cdot{\bf b}.
\end{align}
Alternatively, we can also use $M_\o^\cD$, which does not involve vacuum shift to give \footnote{Though we study vortical and drift perturbations individually at $O(\pd)$, the generation of charge density in vortical solution and generation of heat current in drift solution are connected by Onsager relation \cite{Bu:2019qmd}.}
\begin{align}\label{J0cD}
  \D J^0&=-2(2p_{,B^2}+M_{\o,\m}^\cD)B\o =\(\frac{B}{4\p^2}+\frac{\c}{\p^2}\)\bo\cdot{\bf b}.
\end{align}
We can easily convince ourselves that the structure of the solution $j_{(1)\cV}^\m\propto (u+b)^\m$ dictates that
\begin{align}\label{final_J5}
  \D {\bf J}_A=\(\frac{B}{4\p^2}+\frac{\c}{\p^2}\){\boldsymbol\omega}.
\end{align}
The vacuum parts of \eqref{final_J} and \eqref{final_J5} are in agreement with \cite{Hattori:2016njk}. The medium part for \eqref{final_J5} is consistent with the standard CVE result.

Let us further work out the frame transformation that connects \eqref{cons_sol_1} with \eqref{cons_sol}. The frame transformation amounts to a redefinition of temperature and chemical potential $T\to T+\d T$ and $\m\to\m+\d\m$, giving the following matching equations, 
\begin{align}\label{frm_trs}
  &f_\cN=f_\cN'+\d T\,n_{,{}_T}+\d \m\,n_{,\m},\no
  &f_\cE=f_\cE'+\d T\,\e_{,{}_T}+\d \m\,\e_{,\m},\no
  &f_\cP=f_\cP'+\d T\,\P_{,{}_T}+\d \m\,\P_{,\m},\no
  &f_\cT={f'}_\cT+\frac{B^2}{3}\(\d T\,\a_{{}_{BB,T}}+\d\m\,\a_{{}_{BB},\m} \),
\end{align}
Now we can plug $M_\o^{\cV}$ into \eqref{frm_trs} to get
\begin{align}
  \d T\,n_{,{}_T}+\d\m\,n_{,\m}=&-\frac{7B\o}{12\p^2},\no
  \d T\,p_{,{}_T}+\d\m\,p_{,\m}=&-\frac{\m B\o}{12\p^2}.
\end{align}
The equations can be solved by
\begin{align}
  \d T=\frac{3\o\m}{\p^2 T},\qquad\d\m=&-\frac{7\o}{6}.
\end{align}
We can now translate the frame transformation back to the following $F_4$,
\begin{align}
  F_4=\d\(p\cdot(u+b)\)\(\frac{\pd f(p\cdot u)}{\pd T}\d T+\frac{\pd f(p\cdot u)}{\pd\m}\d\m\)e^{\frac{p_T^2}{B}}.
\end{align}
It is instructive to write down the difference of the final solution with $F_4$ added and the vacuum solution,
\begin{align}\label{vortical_comp}
  j_{(1)\cV }^{\m }-j_{{}_\text{vac}}^\m=&(u+b)^{\m}\bigg[ -\frac{\o}{3}\(\frac{p_T^2}{B}+1\)\d'\(p\cdot (u+b)\)f(p\cdot u)\no
  &+\(\frac{2\o p_T^2}{3B}+\frac{\o}{6}-\frac{3\o\m}{2\p^2T^2}(p\cdot u-\m)\)\d\(p\cdot (u+b)\)f'(p\cdot u) \no
  &-\frac{2\o p_T^2}{B^2}p\cdot u\,\d\(p\cdot (u+b)\)f(p\cdot u)\bigg]e^{\frac{p_T^2}{B}}+\frac{\o p_T^\m}{B}\d\(p\cdot (u+b)\)f(p\cdot u)e^{\frac{p_T^2}{B}}.
\end{align}
The structure of \eqref{vortical_comp} suggests the following interpretation: the first line is modification of dispersion, which does not contribute to charge density upon momenta integration. The terms proportional to $\frac{\o}{6}$ and $\frac{3\o\m}{2\p^2T^2}(\m-p\cdot u)$ come from relative shifts of chemical potential $\d\m-\d\m_{{}_\text{vac}}$ and temperature $\d T$ respectively. Because the zeroth order charge density $n$ is independent of temperature, contribution to charge density from temperature shift $\d T\,n_{,T}$ vanishes. The remaining terms with factors of $p_T$ and $p_T^2$ come from deformation of wave function of the LLL states averaged over the fluid cell, which can be interpreted as excitations of LLL states. Note that higher Landau levels are not excited because they are gapped by $\sqrt{B}\gg\o$. One may ask whether the vacuum and medium contributions can be traced back to shifts and excitations respectively. In fact, it is not true. The vacuum part is given by the terms $\propto\frac{2\o p_T^2}{3B}+\frac{\o}{6}$, which is a mixture of excitations and shifts. The medium part indeed comes from excitations. Finally we remark that we cannot naively take the vacuum limit $T\to0$ in our solution because hydrodynamic description breaks down before the limit is reached.

\section{Summary}\label{sec_summary}

We have obtained covariant chiral kinetic theory with Landau level basis. We have used it to study the magnetized plasma subject to transverse electric field. The solution of the Wigner function is the same as the equilibrium one but with a drift velocity just as in system consisting of free fermions. It gives rise to Hall current and heat current.

We have also studied the Wigner function corresponding to a magnetized plasma with a steady vorticity. The resulting solution contains shifts of temperature and chemical potential as well as excitations of the LLL states. It also gives rise to an vector charge density and axial current density. The vacuum parts of both agree with previous studies and medium part of axial current density is consistent with standard CVE result. We find the vacuum contribution comes from the combination of the two effects, while the medium contribution comes from the excitation effect alone.

The current and stress tensor in both cases have been matched to constitutive relations of MHD, allowing us to determine several thermodynamic functions. An apparent discrepancy in the resulting thermodynamic function has been found. The resolution leads to the conclusion that the vacuum state is shifted as the vorticity is turned on adiabatically. The interpretation is in agreement with \cite{Hattori:2016njk}.

The expectation that axial current comes solely from LLL states seems to indicate \eqref{final_J5} is exact to $O(\o)$. Indeed a same result for charge density is obtained for weak magnetic and vorticity fields based on conventional CKT \cite{Yang:2020mtz}. The numerical agreement of \eqref{final_J} and \eqref{final_J5} follows from an emergent symmetry in LLL approximation. We expect the vector charge density to receive corrections from high Landau levels in general. It would be interesting to extend the present work to include higher Landau levels.

Last but not least, our study is based on collisionless kinetic theory. There have been indications that vorticity can induce spin rotation of fermions through collision effect, which could lead to current generation for fermions with anisotropic distribution \cite{Hou:2020mqp}. It is curious to see whether similar mechanism is manifested with Landau level states. We leave it for future studies.

\begin{acknowledgments}
We thank Han Gao for collaboration at early stage of the work. We are grateful to Jianhua Gao, Koichi Hattori and Yi Yin for useful discussions, also to Koichi Hattori and Yi Yin for helpful comments on an early version of the paper. S.L. thanks Aradhya Shukla for collaborations on related works. S.L. is in part supported by NSFC under Grant Nos 12075328, 11735007 and 11675274.

\end{acknowledgments}

\appendix

\section{Projection}

The decomposition of EOM can be easily done by using the following identity
\begin{align}\label{sigma_identity}
  \s^\m{\bar\s}^\n=g^{\m\n}+(g^{\m\r}g^{\n\s}-g^{\m\s}g^{\n\r}-i\e^{\m\n\r\s})n_\r(n_\s n_\l-g_{\s\l})\s^\l
\end{align}
Here $n$ can be viewed as a time-like frame vector. $(n_\s n_\l-g_{\s\l})\s^\l$ is the spatial components of Pauli matrices orthogonal to $n$. Note that \eqref{sigma_identity} splits into an identity part and a Pauli matrix part, both of which contains real and imaginary parts. It is not difficult to see $W(x,p)$ is hermitian from the definition and $({\bar\sigma}^\m)^\dg={\bar\sigma}^\m$. It follows that $j_\m$ is real. Therefore, the splitting gives in total four equations
\begin{align}
  \P_{\m} j^{\m}&=0,\label{eom1_A}\\
  \D_{\m} j^{\m}&=0,\label{eom2_A}\\
  \P^{\m} j^{\n} - \P^{\n} j^{\m}&=-\frac{1}{2} \e^{\m\n\r\s} \D_{\r} j_\s,\label{eom3_A}\\
  \frac{1}{2}\(\D^{\m} j^{\n} - \D^{\n} j^{\m}\)&=\e^{\m\n\r\s} \P_{\r} j_{\s},\label{eom4_A}.
\end{align}
In fact \eqref{eom4_A} is equivalent to \eqref{eom3_A}.

The case of left-handed fermions is similar. The only difference is that we replace $\s$ in the covariant derivative \eqref{cov_der} by ${\bar\s}$ and $\bar{\s}$ in the decomposition \eqref{right_decomp} by $\s$, which simply flips the sign of RHS of \eqref{eom3_A} and \eqref{eom4_A}.

\section{Momenta Calculus}

We show the momenta differentiation and integration in detail with $q^\m=\L_\n^\m p^\n$ as independent variables and
\begin{align}
  \L_\n^\m=\left(\begin{array}{c}
      u\\
      -w_i\\
      -b\\
      \end{array}\right),\qquad\qquad
  {\L^{{}^{-1}}}_\n^{\m}=\left(\begin{array}{c}
      u\\
      w_i\\
      b\\
      \end{array}\right)^T,
\end{align}
being Lorentz transformations between $p^\m$ and $q^\m$ where $u^\m,\,w_{i=1,2}^\m$ and $b^\m$ are basis row vectors which are orthogonal to one another and normalized as $u^2=1,\,w_i^2=b^2=-1$. The metric in most minus signature can be written as $g^{\m\n}=u^\m u^\n-w_i^\m w_i^\n-b^\m b^\n$ with summation over transverse index $i=1,2$. Then $p^2=(p\cdot u)^2-(p\cdot w_i)(p\cdot w_i)-(p\cdot b)^2\equiv q_0^2-q_T^2-q_3^2=q^2$ with $q_0\equiv p\cdot u,\,q_i\equiv-p\cdot w_i,\,q_3\equiv-p\cdot b$, which gives $p_T^\m=q_i w_i^\m$ and $p_T^2=-q_T^2$.

Note there are gradients in $q_0=p\cdot u$ and $q_T^2=-p_T^2$ since they depend on $u^\m$. Explicitly, $\pd_\m(p\cdot u)=p^\l\pd_\m u_\l$ and $\pd_\m p_T^2=\pd_\m\(p^2-(p\cdot u)^2+(p\cdot b)^2\)=-2p\cdot u\,p^\l\pd_\m u_\l$. Moreover, $p_T^\m,\,E^\m$ and $u_{(1)}^\m$ lie in the transverse plane spanned by $w_i^\m$ and therefore give vanishing dot products with $u^\m$ and $b^\m$. We repeatedly use these properties in the text and the following calculations.

The momenta differentiation can be interpreted as
\begin{align}\label{chain}
  \frac{\pd }{\pd p^{\m }}=\frac{\pd }{\pd(p\cdot u)}u_{\m }+\frac{\pd }{\pd(p\cdot b)}b_{\m }+\frac{\pd }{\pd(p\cdot w_i)}w_{\m }^i.
\end{align}
Multiplying it by $u^\m$ and $b^\m$, one gets $u^{\m }\frac{\pd }{\pd p^{\m }}=\frac{\pd }{\pd(p\cdot u)}$ and $b^{\m }\frac{\pd }{\pd p^{\m }}=-\frac{\pd }{\pd(p\cdot b)}$.
We can then write the operator $\hP_\m\equiv \P_\m-p_\m=-\frac{1}{12}(\pd_\l^p \pd_x^\l)\frac{\pd}{\pd p_\n}F_{\m\n}$ explicitly as follows. Using $\pd ^{\l }u^{\r }=-\o \e^{\l \r \a \b } b_{\a } u_{\b }$, we get
\begin{align}
  \hP_\m=&\frac{B \o}{12}\frac{\pd }{\pd p_{\l }}\frac{\pd }{\pd p_{\n }}\(b_{\l } b_{\m } u_{\n }-b_{\l } b_{\n } u_{\m }+g_{\l \m } u_{\n }-g_{\l \n } u_{\m }\) \no
  =&\frac{B \o}{12}\[\frac{\pd }{\pd(p\cdot u)}\(\frac{\pd }{\pd p^{\m }}-\frac{\pd }{\pd(p\cdot b)}b_{\m }\)-\frac{\pd }{\pd p_{\l }}\frac{\pd }{\pd p_{\n }}\(g_{\l \n }+b_{\l } b_{\n }\) u_{\m }\].
\end{align}
Applying the chain rule \eqref{chain} in the first part, we obtain
\begin{align}
  \hP_\m=&\frac{B \o}{12}\[\frac{\pd }{\pd(p\cdot u)}\(\frac{\pd }{\pd(p\cdot u)}u_{\m }+\frac{\pd }{\pd(p\cdot w_i)}w_{\m }^i\)-\frac{\pd }{\pd p_{\l }}\frac{\pd }{\pd p_{\n }}\(g_{\l \n }+b_{\l } b_{\n }\) u_{\m }\]\no
  =&\frac{B \o}{12}\[\frac{\pd }{\pd(p\cdot u)}\frac{\pd }{\pd(p\cdot w_i)}w_{\m }^i-\frac{\pd }{\pd p_{\l }}\frac{\pd }{\pd p_{\n }}\(g_{\l \n }+b_{\l } b_{\n }-u_{\l } u_{\n }\) u_{\m }\]\no
  =&\frac{B \o}{12}\(\frac{2p_\m^T }{B}\frac{\pd }{\pd(p\cdot u)}+u_{\m }P_{\l\n} \frac{\pd }{\pd p_{\l }}\frac{\pd }{\pd p_{\n }}\),
\end{align}
where, in the last equality, we have identified the operator $\frac{\pd }{\pd(p\cdot w_i)}w_{\m }^i=\frac{2p_\m^T }{B}$ upon acting on functions depending on $p_T^\m$ through $e^{ \frac{p_T^2}{B} }$ only. Also, we have let $\frac{\pd }{\pd(p\cdot u)}$ pass over $\frac{2p_\m^T }{B}$ since $p_\m^T$ is independent of $p\cdot u$.

When solving the kinetic equations, we have repeatedly used the vanishing integration by parts
\begin{align}
  &p\cdot(u+b) \d'\(p\cdot(u+b)\)+\d(p\cdot(u+b))=0,\no
  &\text{or} \quad \(q_0-q_3\) \d'\(q_0-q_3\)+\d(q_0-q_3)=0,
\end{align}

To include contribution from left-handed fermions, we generalize \eqref{cov_sol} and \eqref{distr} as
\begin{align}
  j_s=&\d(q_0-sq_3)f_s(q_0)e^{ \frac{-q_T^2}{B} }, \no
  f_s(q_0)=&\frac{2}{(2\p)^3}\sum_{r=\pm}\frac{ r \th(rq_0)}{e^{r \(q_0-\m_s\)/T}+1},
\end{align}
where helicity $s=\pm$ with $\m_+=\m_{{}_R}$ and $\m_-=\m_{{}_L}$ respectively. With shorthand notations $\int\equiv\int dq_0 dq_3,\,\d_s\equiv\d(q_0-sq_3)$ and $f_s\equiv f_s(q_0)$, the following integrals are useful to perform momenta integration
\begin{align}
  \int \(\d_s f_s\)'=-1,\quad \int \d_s f_s'=-1,\quad \to \quad \int \d_s'f_s=0,
\end{align}
\begin{align}
  &\int \(q_0\d_s f_s\)'=0,\quad \int \d_s f_s=\m_s,\quad \to \quad \int \,q_0 \(\d_s f_s\)'=-\m_s,\no
  \int\,q_0 \d_sf_s'=-\m_s,&\quad \to \quad \int\,q_0 \d_s'f_s=0,\quad \int \,sq_3 \d_s f_s'=-\m_s,\quad \to \quad \int \,sq_3 \d_s' f_s=\m_s,
\end{align}
\begin{align}
  \int \,q_0\d_s f_s=\int \,sq_3\d_s f_s=\frac{\m_s^2}{2}+\frac{\p^2 T^2}{6},
\end{align}
\begin{align}
  \int \, q_0^2\d_s f_s=\int \, sq_3 q_0\d_s f_s=\frac{\m_s}{3} \(\m_s^2+\p^2 T^2\),
\end{align}
\begin{align}
  &\int dq_1 dq_2\,e^{\frac{-q_T^2}{B}}=\int dq_1 dq_2\frac{q_T^2}{B}e^{\frac{-q_T^2}{B}}=2\int dq_1 dq_2\frac{q_i^2}{B}e^{\frac{-q_T^2}{B}}=\p B.
\end{align}
Then, for the $O(1)$ solution, we have
\begin{align}
  &u_\m J_{(0)}^\m=\int d^4p\,u_\m j_{(0)}^\m=\int d^4q\sum_{s=\pm}j_s=\frac{\m B}{2\p^2},\no
  &P^{\m \n} J_\n^{(0)}=\int d^4p P^{\m \n} j_\n^{(0)}=0,\no
  &b^\m b^\n J_\n^{(0)}=\int d^4p\,b^\m b^\n j_\n^{(0)}=-\int d^4q\sum_{s=\pm}sb^\m j_s=-\frac{\m_5 B}{2\p^2}b^\m,
\end{align}
\begin{align}
  &u_\m u_\n T_{(0)}^{\m\n}=\int d^4p\,u_\m u_\n \frac{1}{2} p^{\{\m} j_{(0)}^{\n\}}=\int d^4q\,q_0\sum_{s=\pm}j_s=\frac{\c_{{}_V} B}{2\p^2},\no
  &P_{\m \n} T_{(0)}^{\m\n}=\int d^4p P_{\m \n} \frac{1}{2} p^{\{\m} j_{(0)}^{\n\}}=0,\no
  &b_\m b_\n T_{(0)}^{\m\n}=\int d^4p\,b_\m b_\n \frac{1}{2} p^{\{\m} j_{(0)}^{\n\}}=\int d^4q\sum_{s=\pm}sq_3 \,j_s=\frac{\c_{{}_V} B}{2\p^2},
\end{align}
\begin{align}
  &\D^{\m \a}\D^{\n \b}T_{\a\b}^{(0)}=\int d^4p\,b^\m b^\a b^\n b^\b \frac{1}{2}p_{\{\a} j_{\b\}}^{(0)}=\frac{\o}{B}\int d^4q\sum_{s=\pm}b^\m b^\n\,sq_3 j_s=\frac{\o}{2\p^2} \c_{{}_V}b^\m b^\n,
\end{align}
\begin{align}
  P^{\m \a}u^\b T_{\a\b}^{(0)}=&\int d^4p P^{\m \a} u^\b p_{\{\a} j_{\b\}}^{(0)}=0,\no
  b^{\m}b^{\a}u^\b T_{\a\b}^{(0)}=&\int d^4p\,b^\m b^\a u^\b \frac{1}{2} p_{\{\a} j_{\b\}}^{(0)}=-b^\m\int d^4q\sum_{s=\pm}\(q_3+sq_0\)j_s=\frac{-\c_{{}_A}}{2\p^2}b^\m,
\end{align}
which give
\begin{align}
  &\cN_{(0)}=\frac{\m B}{2\p^2},\qquad
  \cJ_{(0)}^\m=-\D^{\m\n} J_\n^{(0)}=\frac{\m_5 B}{2\p^2}b^\m,\no
  &\cE_{(0)}=\frac{\c_{{}_V} B}{2\p^2},\qquad
  \cP_{(0)}=\frac{1}{3}\D_{\m \n} T_{(0)}^{\m\n}=\frac{\c_{{}_V} B}{6\p^2},\qquad
  \cQ_{(0)}^\m=\D^{\m \a}u^\b T_{\a\b}^{(0)}=-\frac{\c_{{}_A}}{2\p^2}b^\m,\no
  &\cT_{(0)}^{\m\n}=\frac{1}{2}\(\D^{\m \a}\D^{\n \b}+\D^{\n \a}\D^{\m \b}-\frac{2}{3}\D^{\m \n}\D^{\a \b}\)T_{\a\b}^{(0)}=\frac{\c_{{}_V} B}{6\p^2}\(2b^\m b^\n-P^{\m \n}\).
\end{align}
The following are $O(\pd)$ solutions. Firstly, for drift solution, the nontrivial components are
\begin{align}
\cJ_{(1)\cD }^{\m }=&-\D^{\m\n}J_\n^{(1)\cD }=-\int d^4p\,P^{\m\n}j_\n^{(1)\cD }=-\int d^4p\,P^{\m\n}\sum_{s=\pm}u_{\n}^{(1)}j_s\no
  =&\int d^4q\,u_{(1)}^{\m}\sum_{s=\pm}j_s=\frac{\m }{4\p^2} \e^{\m \n \r \s } f_{\n \r } b_{\s }=-\frac{\m }{2\p^2} \e^{\m \n \r \s } u_{\n } E_{\r } b_{\s },
\end{align}
\begin{align}
  \cQ_{(1)\cD }^\m=&-\D^{\m \a } u^{\b } T_{\a \b }^{(1)\cD }=-\int d^4p\,P^{\m \a } u^{\b }\frac{1}{2}p_{\{\a}j_{\b\}}^{(1)\cD }\no
  =&\frac{1}{2}\int d^4p\,\sum_{s=\pm}\[p\cdot u_{(1)}p_T^{\m }\(\frac{\pd }{\pd(p\cdot u)}-\frac{2p\cdot u}{B}\)j_s+ p\cdot u\,u_{(1)}^{\m }j_s\]\no
  =&\frac{1}{2}\int d^4q\,u_{(1)}^{\m }\sum_{s=\pm}\[\frac{q_T^2}{2}\(\frac{2q_0}{B}-\frac{\pd }{\pd q_0}\) j_s+q_0 j_s\]\no
  =&\frac{1}{2\p^2}\(\frac{B}{4}+\chi_{{}_V} \)u_{(1)}^{\m }\no
  =&-\frac{1}{2\p^2}\(\frac{B}{4}+\chi_{{}_V} \)\e^{\m \n \r \s }u_{\n } E_{\r } b_{\s }.
\end{align}
All the other components from drift solution are vanishing upon integration over $q_i$ odd functions.
For vortical solution, we have
\begin{align}
  &u_\m J_{(1)\cC }^\m=\int d^4p\,u_\m j_{(1)\cC }^\m=\frac{\o }{B}\int d^4q\sum_{s=\pm}q_T^2\(\frac{2q_0}{B}j_s-j_s'+\frac{1}{3}\d_s f_s' e^{\frac{-q_T^2}{B}}\)=\frac{\o}{2\p^2}\(2\c_{{}_V}+\frac{2}{3}B\),\no
  &P^{\m\n}J_\n^{(1)\cC }=\int d^4p P^{\m\n} j_\n^{(1)\cC }=0,\no
  &b^\m b^\n J_\n^{(1)\cC }=\int d^4p \,b^\m b^\n j_\n^{(1)\cC }=-\frac{\o }{B}b^\m\int d^4q\sum_{s=\pm}sq_T^2\(\frac{2q_0}{B}j_s-j_s'+\frac{1}{3}\d_s f_s'e^{\frac{-q_T^2}{B}}\)\no
  &\qquad\qquad=-\frac{\o}{2\p^2}2\c_{{}_A}b^\m,
\end{align}
\begin{align}
  &u_\m u_\n T_{(1)\cC }^{\m\n}=\int d^4p\,u_\m u_\n\frac{1}{2}p^{\{\m} j_{(1)\cC }^{\n\}}
  =\frac{\o }{B}\int d^4q\sum_{s=\pm}q_0 q_T^2\(\frac{2q_0}{B}j_s-j_s'+\frac{1}{3}\d_s f_s' e^{\frac{-q_T^2}{B}}\)\no
  &\qquad\qquad=\frac{\o}{2\p^2}\(2\x_{{}_V}+\frac{2}{3}\m B\),\no
  &P_{\m \n}T_{(1)\cC }^{\m\n}=\int d^4p P_{\m \n}\frac{1}{2}p^{\{\m} j_{(1)\cC }^{\n\}}=\frac{\o}{B}\int d^4q\sum_{s=\pm}q_T^2 j_s=\frac{\o}{2\p^2}\m B,\no
  &b_\m b_\n T_{(1)\cC }^{\m\n}=\int d^4p\,b_\m b_\n \frac{1}{2}p^{\{\m} j_{(1)\cC }^{\n\}}=\frac{\o}{B}\int d^4q\sum_{s=\pm}sq_3 q_T^2\(\frac{2q_0}{B}j_s-j_s'+\frac{1}{3}\d_s f_s' e^{\frac{-q_T^2}{B}}\)\no
  &\qquad\qquad=\frac{\o}{2\p^2}\(2\x_{{}_V}-\frac{1}{3}\m B\),
\end{align}
\begin{align}
  &\D^{\m \a}\D^{\n \b}T_{\a\b}^{(1)\cC }=\int d^4p\(P^{\m\a}P^{\n\b}+b^\m b^\a b^\n b^\b \)\frac{1}{2}p_{\{\a} j_{\b\}}^{(1)\cC }\no
  &=\frac{\o}{B}\int d^4p\sum_{s=\pm}\[p_T^{\m}p_T^{\n} j_s+b^\m b^\n \(-sp\cdot b\) p_T^2\(\frac{\pd j_s}{\pd(p\cdot u)}-\frac{1}{3}\d_s f_s e^{\frac{p_T^2}{B}}-\frac{2p\cdot u}{B}j_s\)\]\no
  &=\frac{\o}{B}\int d^4q\sum_{s=\pm}\[\frac{1}{2}P^{\m\n}q_T^2 j_s+b^\m b^\n\,sq_3 q_T^2\(\frac{2q_0}{B}j_s-j_s'+\frac{1}{3}\d_s f_s' e^{\frac{-q_T^2}{B}}\)\]\no
  &=\frac{1}{2}P^{\m\n}\frac{\o}{2\p^2}\m B+b^\m b^\n\frac{\o}{2\p^2}\(2\x_{{}_V}-\frac{1}{3}\m B\),
\end{align}
\begin{align}
  &P^{\m \a} u^\b T_{\a\b}^{(1)\cC }=\int d^4p P^{\m \a} u^\b \frac{1}{2}p_{\{\a} j_{\b\}}^{(1)\cC }=0,\no
  &b^\m b^\a u^\b T_{\a\b}^{(1)\cC }=\int d^4p\,b^{\m}b^{\a} u^\b \frac{1}{2}p_{\{\a} j_{\b\}}^{(1)\cC }=-\frac{b^\m\o}{2B}\int d^4q\sum_{s=\pm}\(q_3+sq_0\)\no
  &\qquad\times q_T^2\(\frac{2q_0}{B}j_s-j_s'+\frac{1}{3}\d_s f_s' e^{\frac{-q_T^2}{B}}\)=-\frac{\o}{2\p^2}\(2\x_{{}_A}+\frac{1}{6}\m_5 B\)b^\m.
\end{align}
The $j_{(1)\cA }^{\m}$ part does not contribute to the final vortical result since we have
\begin{align}
\tP e^{\frac{p_T^2}{B}}\sim P_{\l\n} \frac{\pd }{\pd p_{\l }}\frac{\pd }{\pd p_{\n }}e^{\frac{p_T^2}{B}}=\frac{\pd }{\pd q_i}\frac{\pd }{\pd q_i}e^{\frac{-q_T^2}{B}}=\frac{4}{B}\(\frac{q_T^2}{B}-1\)e^{\frac{-q_T^2}{B}},
\end{align}
which gives vanishing integral by noting $\int dq_1 dq_2\,e^{\frac{-q_T^2}{B}}=\int dq_1 dq_2\frac{q_T^2}{B}e^{\frac{-q_T^2}{B}}$. Thus
\begin{align}
  &\cN_{(1)\cV }=\frac{\o}{2\p^2}\(2\c_{{}_V}+\frac{2}{3}B\),\qquad \cJ_{(1)\cV }^\m=-\D^{\m\n}J_\n^{(1)\cV }=\frac{\o}{2\p^2}2\c_{{}_A}b^\m,\no
  &\cE_{(1)\cV }=\frac{\o}{2\p^2}\(2\x_{{}_V}+\frac{2}{3}\m B\),\qquad \cP_{(1)\cV }=\D_{\m\n}T_{(1)\cV }^{\m\n}=\frac{\o}{6\p^2}\(2\x_{{}_V}+\frac{2}{3}\m B\),\no
  &\cQ_{(1)\cV }^\m=-\D^{\m \a} u^\b T_{\a\b}^{(1)\cV }=\frac{\o}{2\p^2}\(2\x_{{}_A}+\frac{1}{6}\m_5 B\)b^\m,\no
  &\cT_{(1)\cV }^{\m\n}=\frac{1}{2}\(\D^{\m \a}\D^{\n \b}+\D^{\n \a}\D^{\m \b}-\frac{2}{3}\D^{\m \n}\D^{\a \b}\)T_{\a\b}^{(1)\cV }\no
  &\qquad=\frac{\o}{6\p^2}\(2\x_{{}_V}-\frac{5}{6}\m B\)\(2b^\m b^\n-P^{\m \n}\).
\end{align}

\section{Useful Formulas}

We have repeatedly used the contraction formulas of two anti-symmetric tensors,
\begin{align}
  \e^{\m \n \r \s } \e_{\m \n \a \b }=-2\left|\begin{array}{cc}
      \d_\a^\r & \d_\b^\r\\
      \d_\a^\s & \d_\b^\s\\
      \end{array}\right|,\no
  \e^{\m \n \r \s } \e_{\m \l \a \b }=-\left|\begin{array}{ccc}
      \d_\l^\n & \d_\a^\n & \d_\b^\n\\
      \d_\l^\r & \d_\a^\r & \d_\b^\r\\
      \d_\l^\s & \d_\a^\s & \d_\b^\s\\
      \end{array}\right|.
\end{align}
By repeated use of the five-index cyclic identity,
\begin{align}
  \e^{\m \n \r \s } p^\l+\e^{\n \r \s \l } p^\m+\e^{\r \s \l \m } p^\n+\e^{\s \l \m \n } p^\r+\e^{\l \m \n \r } p^\s=0,
\end{align}
we can prove the identities in \eqref{identity}. For the first one,
\begin{align}
  \e^{\m \n \r \s } f_{\r \l } (u+b)_{\s } u^{\l }=-\(\e^{\n \r \s \l } u^\m+\e^{\r \s \l \m } u^\n+\e^{\s \l \m \n } u^\r+\e^{\l \m \n \r } u^\s\) f_{\r \l } (u+b)_{\s }.
\end{align}
Upon moving the third term $\e^{\s \l \m \n } u^\r$ of the RHS to the LHS, one finds the LHS doubles. At the same time, the $u_\s$ parts in the first and second terms vanish by noting $f_{\r\l}=E_\r u_\l-E_\l u_\r$ and the anti-symmetry of $\e^{\n \r \s \l }$ and $\e^{\r \s \l \m }$. Rearranging the indices, one gets
\begin{align}
  2\e^{\m \n \r \s } f_{\r \l } (u+b)_{\s } u^{\l }=u^{[\m}\e^{\n] \r \l \s }f_{\r \l }b_\s+\e^{\m \n \r \l } f_{\r \l }.
\end{align}
The $\e^{\m \n \r \l } f_{\r \l }$ term can be written into $b^{[\m}\e^{\n] \r \s \l }f_{\r \l } b_{\s }$ by starting from
\begin{align}
  b^{\m } \e^{\n \r \l \s } f_{\r \l } b_{\s }=-\(b^\n \e^{\r \l \s \m }+b^\r \e^{\l \s \m \n }+b^\l \e^{\s \m \n \r }+b^{\s } \e^{\m \n \r \l }\) f_{\r \l } b_{\s },
\end{align}
where we can move the first term $b^\n \e^{\r \l \s \m}$ of the RHS to the LHS to produce $b^{[\m}\e^{\n] \r \l \s } f_{\r \l } b_{\s }$ in the LHS. Then by noting $b^\r f_{\r \l }=b^\l f_{\r \l }=0$ in the second and third terms, we do get $\e^{\m \n \r \l } f_{\r \l }=b^{[\m}\e^{\n] \r \l \s }f_{\r \l } b_{\s }$ which gives
\begin{align}
  2\e^{\m \n \r \s } f_{\r \l } (u+b)_{\s } u^{\l }=(u+b)^{[\m}\e^{\n] \r \l \s }f_{\r \l } b_{\s }=2B (u+b)^{[\m}u_{(1)}^{\n]}.
\end{align}
Similarly, for the second identity in \eqref{identity}, one has
\begin{align}
  &\e^{\m \n \r \s } f_{\r \l } (u+b)_{\s } p_T^{\l }=-\(\e^{\n \r \s \l } p_T^\m+\e^{\r \s \l \m } p_T^\n+\e^{\s \l \m \n } p_T^\r+\e^{\l \m \n \r } p_T^\s\) f_{\r \l } (u+b)_{\s }\no
  &\to\quad 2\e^{\m \n \r \s } f_{\r \l } (u+b)_{\s } p_T^{\l }=p_T^{[\m} \e^{\n] \r \l \s } f_{\r \l }b_\s=2p_T^{[\m}u_{(1)}^{\n] }.
\end{align}

\bibliographystyle{unsrt}
\bibliography{magneto-vortical.bib}

\begin{thebibliography}{10}

\bibitem{Vilenkin:1980fu}
A.~Vilenkin.
\newblock {EQUILIBRIUM PARITY VIOLATING CURRENT IN A MAGNETIC FIELD}.
\newblock {\em Phys. Rev. D}, 22:3080--3084, 1980.

\bibitem{Kharzeev:2004ey}
Dmitri Kharzeev.
\newblock {Parity violation in hot QCD: Why it can happen, and how to look for
  it}.
\newblock {\em Phys. Lett. B}, 633:260--264, 2006.

\bibitem{Kharzeev:2007tn}
D.~Kharzeev and A.~Zhitnitsky.
\newblock {Charge separation induced by P-odd bubbles in QCD matter}.
\newblock {\em Nucl. Phys. A}, 797:67--79, 2007.

\bibitem{Fukushima:2008xe}
Kenji Fukushima, Dmitri~E. Kharzeev, and Harmen~J. Warringa.
\newblock {The Chiral Magnetic Effect}.
\newblock {\em Phys. Rev. D}, 78:074033, 2008.

\bibitem{Son:2009tf}
Dam~T. Son and Piotr Surowka.
\newblock {Hydrodynamics with Triangle Anomalies}.
\newblock {\em Phys. Rev. Lett.}, 103:191601, 2009.

\bibitem{Neiman:2010zi}
Yasha Neiman and Yaron Oz.
\newblock {Relativistic Hydrodynamics with General Anomalous Charges}.
\newblock {\em JHEP}, 03:023, 2011.

\bibitem{Vilenkin:1980zv}
A.~Vilenkin.
\newblock {QUANTUM FIELD THEORY AT FINITE TEMPERATURE IN A ROTATING SYSTEM}.
\newblock {\em Phys. Rev. D}, 21:2260--2269, 1980.

\bibitem{Erdmenger:2008rm}
Johanna Erdmenger, Michael Haack, Matthias Kaminski, and Amos Yarom.
\newblock {Fluid dynamics of R-charged black holes}.
\newblock {\em JHEP}, 01:055, 2009.

\bibitem{Banerjee:2008th}
Nabamita Banerjee, Jyotirmoy Bhattacharya, Sayantani Bhattacharyya, Suvankar
  Dutta, R.~Loganayagam, and P.~Surowka.
\newblock {Hydrodynamics from charged black branes}.
\newblock {\em JHEP}, 01:094, 2011.

\bibitem{Landsteiner:2011cp}
Karl Landsteiner, Eugenio Megias, and Francisco Pena-Benitez.
\newblock {Gravitational Anomaly and Transport}.
\newblock {\em Phys. Rev. Lett.}, 107:021601, 2011.

\bibitem{Hattori:2016njk}
Koichi Hattori and Yi~Yin.
\newblock {Charge redistribution from anomalous magnetovorticity coupling}.
\newblock {\em Phys. Rev. Lett.}, 117(15):152002, 2016.

\bibitem{Liu:2017spl}
Yizhuang Liu and Ismail Zahed.
\newblock {Pion Condensation by Rotation in a Magnetic field}.
\newblock {\em Phys. Rev. Lett.}, 120(3):032001, 2018.

\bibitem{Chen:2015hfc}
Hao-Lei Chen, Kenji Fukushima, Xu-Guang Huang, and Kazuya Mameda.
\newblock {Analogy between rotation and density for Dirac fermions in a
  magnetic field}.
\newblock {\em Phys. Rev. D}, 93(10):104052, 2016.

\bibitem{Cao:2019ctl}
Gaoqing Cao and Lianyi He.
\newblock {Rotation induced charged pion condensation in a strong magnetic
  field: A Nambu\textendash{}Jona-Lasino model study}.
\newblock {\em Phys. Rev. D}, 100(9):094015, 2019.

\bibitem{Chen:2019tcp}
Hao-Lei Chen, Xu-Guang Huang, and Kazuya Mameda.
\newblock {Do charged pions condense in a magnetic field with rotation?}
\newblock 10 2019.

\bibitem{Bu:2019qmd}
Yanyan Bu and Shu Lin.
\newblock {Magneto-vortical effect in strongly coupled plasma}.
\newblock {\em Eur. Phys. J. C}, 80(5):401, 2020.

\bibitem{Fukushima:2020ncb}
Kenji Fukushima, Takuya Shimazaki, and Lingxiao Wang.
\newblock {Mode decomposed chiral magnetic effect and rotating fermions}.
\newblock {\em Phys. Rev. D}, 102(1):014045, 2020.

\bibitem{Kovtun:2016lfw}
Pavel Kovtun.
\newblock {Thermodynamics of polarized relativistic matter}.
\newblock {\em JHEP}, 07:028, 2016.

\bibitem{Hernandez:2017mch}
Juan Hernandez and Pavel Kovtun.
\newblock {Relativistic magnetohydrodynamics}.
\newblock {\em JHEP}, 05:001, 2017.

\bibitem{Grozdanov:2016tdf}
Sa\v{s}o Grozdanov, Diego~M. Hofman, and Nabil Iqbal.
\newblock {Generalized global symmetries and dissipative magnetohydrodynamics}.
\newblock {\em Phys. Rev. D}, 95(9):096003, 2017.

\bibitem{Hongo:2020qpv}
Masaru Hongo and Koichi Hattori.
\newblock {Revisiting relativistic magnetohydrodynamics from quantum
  electrodynamics}.
\newblock 5 2020.

\bibitem{Hattori:2017usa}
Koichi Hattori, Yuji Hirono, Ho-Ung Yee, and Yi~Yin.
\newblock {MagnetoHydrodynamics with chiral anomaly: phases of collective
  excitations and instabilities}.
\newblock {\em Phys. Rev. D}, 100(6):065023, 2019.

\bibitem{Huang:2011dc}
Xu-Guang Huang, Armen Sedrakian, and Dirk~H. Rischke.
\newblock {Kubo formulae for relativistic fluids in strong magnetic fields}.
\newblock {\em Annals Phys.}, 326:3075--3094, 2011.

\bibitem{Finazzo:2016mhm}
Stefano~Ivo Finazzo, Renato Critelli, Romulo Rougemont, and Jorge Noronha.
\newblock {Momentum transport in strongly coupled anisotropic plasmas in the
  presence of strong magnetic fields}.
\newblock {\em Phys. Rev. D}, 94(5):054020, 2016.
\newblock [Erratum: Phys.Rev.D 96, 019903 (2017)].

\bibitem{Lin:2019fqo}
Shu Lin and Lixin Yang.
\newblock {Chiral kinetic theory from Landau level basis}.
\newblock {\em Phys. Rev. D}, 101(3):034006, 2020.

\bibitem{Hattori:2016lqx}
Koichi Hattori, Shiyong Li, Daisuke Satow, and Ho-Ung Yee.
\newblock {Longitudinal Conductivity in Strong Magnetic Field in Perturbative
  QCD: Complete Leading Order}.
\newblock {\em Phys. Rev. D}, 95(7):076008, 2017.

\bibitem{Sheng:2017lfu}
Xin-li Sheng, Dirk~H. Rischke, David Vasak, and Qun Wang.
\newblock {Wigner functions for fermions in strong magnetic fields}.
\newblock {\em Eur. Phys. J. A}, 54(2):21, 2018.

\bibitem{Son:2012wh}
Dam~Thanh Son and Naoki Yamamoto.
\newblock {Berry Curvature, Triangle Anomalies, and the Chiral Magnetic Effect
  in Fermi Liquids}.
\newblock {\em Phys. Rev. Lett.}, 109:181602, 2012.

\bibitem{Son:2012zy}
Dam~Thanh Son and Naoki Yamamoto.
\newblock {Kinetic theory with Berry curvature from quantum field theories}.
\newblock {\em Phys. Rev. D}, 87(8):085016, 2013.

\bibitem{Stephanov:2012ki}
M.~A. Stephanov and Y.~Yin.
\newblock {Chiral Kinetic Theory}.
\newblock {\em Phys. Rev. Lett.}, 109:162001, 2012.

\bibitem{Gao:2012ix}
Jian-Hua Gao, Zuo-Tang Liang, Shi Pu, Qun Wang, and Xin-Nian Wang.
\newblock {Chiral Anomaly and Local Polarization Effect from Quantum Kinetic
  Approach}.
\newblock {\em Phys. Rev. Lett.}, 109:232301, 2012.

\bibitem{Pu:2010as}
Shi Pu, Jian-hua Gao, and Qun Wang.
\newblock {A consistent description of kinetic equation with triangle anomaly}.
\newblock {\em Phys. Rev. D}, 83:094017, 2011.

\bibitem{Chen:2012ca}
Jiunn-Wei Chen, Shi Pu, Qun Wang, and Xin-Nian Wang.
\newblock {Berry Curvature and Four-Dimensional Monopoles in the Relativistic
  Chiral Kinetic Equation}.
\newblock {\em Phys. Rev. Lett.}, 110(26):262301, 2013.

\bibitem{Hidaka:2016yjf}
Yoshimasa Hidaka, Shi Pu, and Di-Lun Yang.
\newblock {Relativistic Chiral Kinetic Theory from Quantum Field Theories}.
\newblock {\em Phys. Rev. D}, 95(9):091901, 2017.

\bibitem{Manuel:2013zaa}
Cristina Manuel and Juan~M. Torres-Rincon.
\newblock {Kinetic theory of chiral relativistic plasmas and energy density of
  their gauge collective excitations}.
\newblock {\em Phys. Rev. D}, 89(9):096002, 2014.

\bibitem{Manuel:2014dza}
Cristina Manuel and Juan~M. Torres-Rincon.
\newblock {Chiral transport equation from the quantum Dirac Hamiltonian and the
  on-shell effective field theory}.
\newblock {\em Phys. Rev. D}, 90(7):076007, 2014.

\bibitem{Wu:2016dam}
Yan Wu, Defu Hou, and Hai-cang Ren.
\newblock {Field theoretic perspectives of the Wigner function formulation of
  the chiral magnetic effect}.
\newblock {\em Phys. Rev. D}, 96(9):096015, 2017.

\bibitem{Mueller:2017arw}
Niklas Mueller and Raju Venugopalan.
\newblock {Worldline construction of a covariant chiral kinetic theory}.
\newblock {\em Phys. Rev. D}, 96(1):016023, 2017.

\bibitem{Mueller:2017lzw}
Niklas Mueller and Raju Venugopalan.
\newblock {The chiral anomaly, Berry's phase and chiral kinetic theory, from
  world-lines in quantum field theory}.
\newblock {\em Phys. Rev. D}, 97(5):051901, 2018.

\bibitem{Huang:2018wdl}
Anping Huang, Shuzhe Shi, Yin Jiang, Jinfeng Liao, and Pengfei Zhuang.
\newblock {Complete and Consistent Chiral Transport from Wigner Function
  Formalism}.
\newblock {\em Phys. Rev. D}, 98(3):036010, 2018.

\bibitem{Gao:2018wmr}
Jian-Hua Gao, Zuo-Tang Liang, Qun Wang, and Xin-Nian Wang.
\newblock {Disentangling covariant Wigner functions for chiral fermions}.
\newblock {\em Phys. Rev. D}, 98(3):036019, 2018.

\bibitem{Carignano:2018gqt}
Stefano Carignano, Cristina Manuel, and Juan~M. Torres-Rincon.
\newblock {Consistent relativistic chiral kinetic theory: A derivation from
  on-shell effective field theory}.
\newblock {\em Phys. Rev. D}, 98(7):076005, 2018.

\bibitem{Lin:2019ytz}
Shu Lin and Aradhya Shukla.
\newblock {Chiral Kinetic Theory from Effective Field Theory Revisited}.
\newblock {\em JHEP}, 06:060, 2019.

\bibitem{Carignano:2019zsh}
Stefano Carignano, Cristina Manuel, and Juan~M. Torres-Rincon.
\newblock {Chiral kinetic theory from the on-shell effective field theory:
  Derivation of collision terms}.
\newblock {\em Phys. Rev. D}, 102(1):016003, 2020.

\bibitem{Liu:2018xip}
Yu-Chen Liu, Lan-Lan Gao, Kazuya Mameda, and Xu-Guang Huang.
\newblock {Chiral kinetic theory in curved spacetime}.
\newblock {\em Phys. Rev. D}, 99(8):085014, 2019.

\bibitem{Weickgenannt:2019dks}
Nora Weickgenannt, Xin-Li Sheng, Enrico Speranza, Qun Wang, and Dirk~H.
  Rischke.
\newblock {Kinetic theory for massive spin-1/2 particles from the
  Wigner-function formalism}.
\newblock {\em Phys. Rev. D}, 100(5):056018, 2019.

\bibitem{Gao:2019znl}
Jian-Hua Gao and Zuo-Tang Liang.
\newblock {Relativistic Quantum Kinetic Theory for Massive Fermions and Spin
  Effects}.
\newblock {\em Phys. Rev. D}, 100(5):056021, 2019.

\bibitem{Hattori:2019ahi}
Koichi Hattori, Yoshimasa Hidaka, and Di-Lun Yang.
\newblock {Axial Kinetic Theory and Spin Transport for Fermions with Arbitrary
  Mass}.
\newblock {\em Phys. Rev. D}, 100(9):096011, 2019.

\bibitem{Wang:2019moi}
Ziyue Wang, Xingyu Guo, Shuzhe Shi, and Pengfei Zhuang.
\newblock {Mass Correction to Chiral Kinetic Equations}.
\newblock {\em Phys. Rev. D}, 100(1):014015, 2019.

\bibitem{Yang:2020hri}
Di-Lun Yang, Koichi Hattori, and Yoshimasa Hidaka.
\newblock {Effective quantum kinetic theory for spin transport of fermions with
  collsional effects}.
\newblock {\em JHEP}, 07:070, 2020.

\bibitem{Liu:2020flb}
Yu-Chen Liu, Kazuya Mameda, and Xu-Guang Huang.
\newblock {Covariant Spin Kinetic Theory I: Collisionless Limit}.
\newblock {\em Chin. Phys. C}, 44(9):094101, 2020.

\bibitem{Hayata:2020sqz}
Tomoya Hayata, Yoshimasa Hidaka, and Kazuya Mameda.
\newblock {Second order chiral kinetic theory under gravity and antiparallel
  charge-energy flow}.
\newblock 12 2020.

\bibitem{chen2021equaltime}
Shile Chen, Ziyue Wang, and Pengfei Zhuang.
\newblock Equal-time kinetic equations in a rotational field, 2021.

\bibitem{Vasak:1987um}
D.~Vasak, M.~Gyulassy, and H.~T. Elze.
\newblock {Quantum Transport Theory for Abelian Plasmas}.
\newblock {\em Annals Phys.}, 173:462--492, 1987.

\bibitem{Elze:1986qd}
H.~T. Elze, M.~Gyulassy, and D.~Vasak.
\newblock {Transport Equations for the \{QCD\} Quark Wigner Operator}.
\newblock {\em Nucl. Phys. B}, 276:706--728, 1986.

\bibitem{Elze:1989un}
Hans-Thomas Elze and Ulrich~W. Heinz.
\newblock {Quark - Gluon Transport Theory}.
\newblock {\em Phys. Rept.}, 183:81--135, 1989.

\bibitem{Zhuang:1995pd}
P.~Zhuang and Ulrich~W. Heinz.
\newblock {Relativistic quantum transport theory for electrodynamics}.
\newblock {\em Annals Phys.}, 245:311--338, 1996.

\bibitem{Note1}
One can equivalently switch between $p^{\mu }$ and $q^{\mu }$ to arrive at the
  same final solution.

\bibitem{Note2}
In making the comparison, we note that \cite {Hernandez:2017mch} uses a
  different signature in metric. Also their definition of electromagnetic field
  or alternatively current differs from ours by a sign. We quote the converted
  constitutive relations of MHD.

\bibitem{Note3}
Here it is more appropriate to regard $b$ as the spin direction of LLL states
  rather than the magnetic field direction.

\bibitem{Kovtun:2012rj}
Pavel Kovtun.
\newblock {Lectures on hydrodynamic fluctuations in relativistic theories}.
\newblock {\em J. Phys. A}, 45:473001, 2012.

\bibitem{Note4}
Though we study vortical and drift perturbations individually at $O({\partial
  })$, the generation of charge density in vortical solution and generation of
  heat current in drift solution are connected by Onsager relation \cite
  {Bu:2019qmd}.

\bibitem{Yang:2020mtz}
Shi-Zheng Yang, Jian-Hua Gao, Zuo-Tang Liang, and Qun Wang.
\newblock {Second-order charge currents and stress tensor in a chiral system}.
\newblock {\em Phys. Rev. D}, 102(11):116024, 2020.

\bibitem{Hou:2020mqp}
Defu Hou and Shu Lin.
\newblock {Polarization Rotation of Chiral Fermions in Vortical Fluid}.
\newblock 8 2020.

\end{thebibliography}

\end{document}